\documentclass[journal]{IEEEtran}

\usepackage{ifpdf}

\usepackage{cite}

\ifCLASSINFOpdf
  \usepackage[pdftex]{graphicx}
\else
  \usepackage[dvips]{graphicx}
\fi
\usepackage{array}
\usepackage{siunitx}
\usepackage{booktabs, makecell, tabularx}
\usepackage{hyperref}
\hypersetup{colorlinks, citecolor=black, linkcolor=black, urlcolor=black}
\usepackage{caption}
\usepackage{booktabs}
\usepackage{subcaption}
\usepackage{xcolor}
\usepackage{comment}
\usepackage{textcomp,mathcomp}
\usepackage{multirow}
\usepackage{multicol}

\usepackage{cuted}
\usepackage{etoolbox}
\AfterEndEnvironment{strip}{\leavevmode}
\usepackage{fancyhdr}
\usepackage{lipsum}
\usepackage{amsmath}
\usepackage{balance}
\usepackage{amssymb}

\usepackage{algorithmic}
\usepackage[top=1.2in,bottom=0.8in,right=1.5cm,left=1.5cm,headheight=30pt]{geometry}

\fancypagestyle{special}{%
\fancyhf{}%
\fancyhead[C]{\copyright {S. J. Babu et al.} {2025}. This is the author's version of the work only for your personal use. The definitive version of Record is under review by ACM, copyright may be transferred without notice}

\setlength{\headheight}{22.41992pt}
}

\begin{document}

\title{Extending Silicon Lifetime: A Review of Design Techniques for Reliable Integrated Circuits}

\author{Shaik~Jani~Babu, Fan~Hu, Linyu~Zhu, Sonal~Singhal,~and~Xinfei~Guo
\thanks{This work is under review by ACM. The work was partially supported by National Science Foundation of China under Grant No. 62201340, and in part by a SJTU Explore-X Research Grant (Corresponding author: Xinfei~Guo).}
\thanks{J.B. Shaik, F. Hu, L. Zhu, and X. Guo are with the University of Michigan – Shanghai
Jiao Tong University Joint Institute, Shanghai Jiao Tong University, Shanghai
200240, China (e-mail: skjanibabu786@sjtu.edu.cn, fan.hu@sjtu.edu.cn, linyuzhu@sjtu.edu.cn, and xinfei.guo@sjtu.edu.cn).}
\thanks{S. Singhal is with Department of Electrical Engineering, Shiv Nadar University, Greater Noida, India (email: sonal.singhal@snu.edu.in).}
\vspace{-20pt}}


\maketitle
\thispagestyle{fancy}
\begin{abstract}
Reliability has become an increasing concern in modern computing. Integrated circuits (ICs) are the backbone of today’s computing devices across industries, including artificial intelligence (AI), consumer electronics, healthcare, automotive, industrial, and aerospace. Moore’s Law has driven the semiconductor IC industry toward smaller dimensions, improved performance, and greater energy efficiency. However, as transistors shrink to atomic scales, aging-related degradation mechanisms such as Bias Temperature Instability (BTI), Hot Carrier Injection (HCI), Time-Dependent Dielectric Breakdown (TDDB), Electromigration (EM), and stochastic aging-induced variations have become major reliability threats. From an application perspective, applications like AI training and autonomous driving require continuous and sustainable operation to minimize recovery costs and enhance safety. Additionally, the high cost of chip replacement and reproduction underscores the need for extended lifespans. These factors highlight the urgency of designing more reliable ICs as key computing infrastructure. This survey addresses the critical issue of aging in ICs, focusing on fundamental degradation mechanisms and mitigation strategies. It provides a comprehensive overview of aging’s impact and the methods to counter it, starting with the root causes of aging and summarizing key monitoring techniques at both circuit and system levels. A detailed analysis of circuit-level mitigation strategies highlights the distinct aging characteristics of digital, analog, and SRAM circuits, emphasizing the need for tailored solutions. The survey also explores emerging software approaches in design automation, aging characterization, and mitigation, which are transforming traditional reliability optimization. Finally, it outlines the challenges and future directions for improving aging management and ensuring the long-term reliability of ICs across diverse applications.
\end{abstract}


%
\IEEEpeerreviewmaketitle


\section{Introduction}
Integrated circuits (ICs) have revolutionized electronic devices since their invention in the late 1950s by Jack Kilby and Robert Noyce ~\cite{kilby1964_ICinv, noyce2007_IC_inv}, transforming consumer electronics, healthcare, transportation, and defense. ICs now power mobile phones, medical implants, autonomous vehicles, high-performance computing clusters, satellite systems, and the Internet of Things (IoT). Emerging fields like quantum computing and neuromorphic systems are poised to further reshape the technological landscape, with silicon chips remaining the essential computing engine. Moore's prediction (known as Moores' law) of a doubling of transistor density every 18 months led to a significant increase in chip performance ~\cite{moore2006_ProgressinDigital, Nigam13_Moore_Rel}. The improvement in performance, driven by continuous advancements in process technology, has resulted in significant developments in VLSI technology, leading to reduced per-unit costs and will continue to be pivotal in the future of technology. Despite their transformative benefits, ICs face significant challenges, including variations in the semiconductor fabrication process and reliability issues cud to reduced device dimensions, higher electric fields, increased power density, and growing chip complexity ~\cite{Diaz2017_CMOSRel, hill2022_CMOSRel_Intro,Guo2018_relIntro,guo2022_ASH_designForRecovery}. Among these challenges, aging-related degradation poses a critical threat to long-term performance and durability, making it a key factor in determining chip lifetime. Chips, like biological organisms such as humans, undergo aging when in operation. Voltage and current stress, high temperatures/voltages, and random variations all contribute to this aging process, similar to how humans experience fatigue under extreme physical conditions ~\cite{guo2020_ICIntro_CircadianRhythmsFuture}. Unfortunately, the key components of chips, transistors and interconnects, both degrade over time, with aging effects worsening at smaller technology nodes and under higher temperatures and voltages. 

Beyond device scaling, aging issues are becoming more pronounced from an application perspective ~\cite{beigi2022reliability}. Chips are now essential in nearly all aspects of daily life, with increasing demand for extended lifespans in various emerging application domains. For example, in the automotive industry, self-driving and intelligent systems rely heavily on thousands of silicon chips, which are expected to last significantly longer than typical consumer electronics like laptops or smartphones ~\cite{angione2022test}. This challenge is complicated by the fact that these expectations assume normal workloads. However, autonomous systems in vehicles often have higher utilization rates, leading to increased circuit temperatures and accelerated aging. Automotive chips are aging much faster than expected in hot climates with sustained high temperatures, raising concerns about the long-term reliability of electrified vehicles and the suitability of advanced-node chips for safety-critical applications. Another demanding domain is artificial intelligence (AI) computing. Training large models often requires hundreds of GPU-years and terabytes of memory, placing extreme demands on device reliability. Since these models are trained in parallel, a single GPU failure can require restarting the entire job ~\cite{emanuele2024iedm}. Even a single uncorrectable DRAM error can affect thousands of GPUs. While recent models like DeepSeek R1 ~\cite{guo2025deepseek} have reduced training times, they still require uninterrupted operation, underscoring the AI industry's growing focus on chip reliability. In summary, these emerging applications are reshaping the computing landscape while highlighting critical concerns about chip reliability. This underscores the importance of understanding the key aging mechanisms, developing effective mitigation strategies, and revisiting design techniques to enhance the longevity and robustness of modern chips. Therefore, it is crucial to revisit aging mechanisms and understand how device-level degradation can lead to system failures. Incorporating aging considerations during or even before the chip design phase is becoming essential for many applications. Traditional flat guardband approaches are no longer sufficient due to increasing demands for higher performance and better energy efficiency. While some aging-aware design techniques are well-researched, others remain in early development stages, highlighting the urgency of this survey. This paper aims to provide a comprehensive overview of how various design techniques address aging mechanisms and how monitoring and mitigation strategies work together. Unlike traditional performance, power, and area (PPA) metrics, aging is still not fully integrated into conventional design automation tools. Therefore, this survey will also review recent aging-aware softwares and design automation algorithms and tools. Additionally, it will examine design techniques developed in recent years that response to the growing demands of the automotive and AI industries.

The rest of this survey is organized as follows. Section ~\ref{backgroud} provides an overview of IC deployment across various industry sectors and discusses related surveys in this domain. Section ~\ref{mechanisms} revisits key device-level aging mechanisms that affect circuit and system performance. Section ~\ref{monitor} examines current aging monitoring techniques used in integrated circuits. Section ~\ref{mitigation} explores aging mitigation approaches at both the circuit and system levels, with circuit-level strategies further categorized into digital circuits, analog circuits, and SRAM with its peripheral circuits. Section ~\ref{software} discusses emerging techniques for aging characterization and mitigation, along with the aging-aware design automation techniques. Section ~\ref{challenge} outlines current challenges and future trends in aging characterization and mitigation. Finally, Section ~\ref{conclusions} concludes the survey.

\begin{figure*}[!t]
\centering
\includegraphics[width=0.8\linewidth]{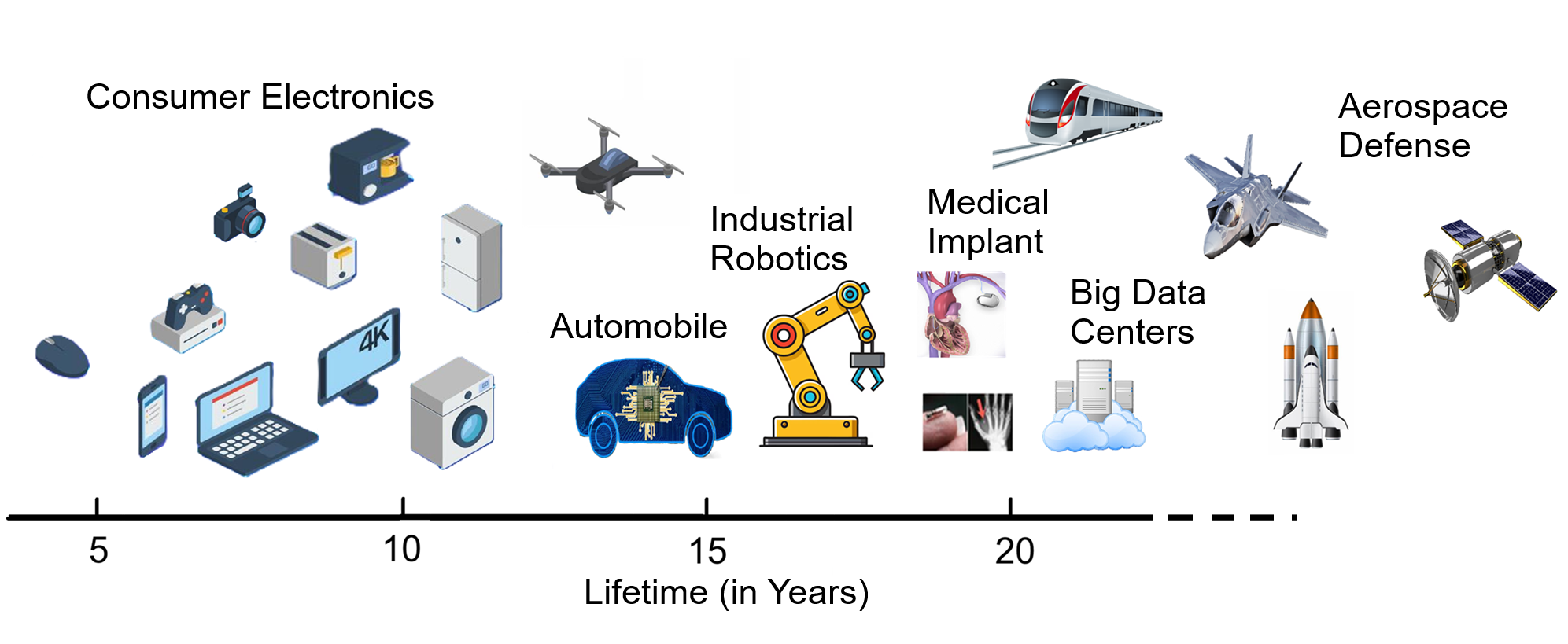}
\caption{Applications of integrated circuits and their expected lifetime.}
\label{IC_applications}
\end{figure*}

\section{Background and Related Surveys}
\label{backgroud}

\pagestyle{special}
This section examines the various applications of ICs and their operational lifespans, highlighting the essential importance of reliability in modern electronic systems. This study examines previous research on integrated circuit reliability, analyzing existing studies and approaches that address issues in this field. This section identifies unresolved concerns in prior works, establishing the motivation for this survey, which aims to address critical research gaps and enhance approaches for improving IC reliability. 

\subsection{Integrated Circuits in Diverse Applications}
Semiconductor chips powered by integrated circuits (ICs) offer advantages in compact size, low power consumption, and high performance, enabling a wide range of applications across major sectors of daily life. The computing landscape has shifted from general-purpose ICs to application-specific and domain-specific chips, where ICs are tailored to specific applications to provide optimal performance and design efficiency for targeted scenarios. This shift has also diversified design requirements, including reliability and lifespan expectations. Figure ~\ref{IC_applications} illustrates six key sectors where integrated circuits play a dominant role in the modern world: (i) Consumer Electronics, (ii) Automotive, (iii) Industrial and Robotics, (iv) Biomedical and Healthcare, Industrial, (v) Computing Infrastructure, and (vi) Aerospace and Defense. An overview of each application is presented below.

\textbf{\textit{Consumer Electronics:}} This industry has expanded significantly over the past three decades, driven by the proliferation of smartphones, tablets, wearable devices, internet of things, and similar technologies. These applications propel progress in computer, communication, and multimedia processing. While lifetime concerns are less critical, extended operation periods and limited cooling support in these devices are raising new reliability challenges ~\cite{Guo2018_relIntro}.

\textbf{\textit{Automotive:}} In the automotive industry, ICs facilitate advanced driver-assistance systems (ADAS), engine management, and in-vehicle infotainment, therefore enhancing safety, performance, and user experiences. The adoption of ADAS in increasing exponentially, driven by the growing demand for electric vehicles (EVs) ~\cite{neumann2024_ICappli_AnalysisADAS}. 
Given the wide range of operation temperatures and mechanical vibrations in autonomous vehicles, IC design engineers must ensure high durability and sustainability across the whole lifespan, which is typically over 15 years.


\textbf{\textit{Industrial and Robotics:}} Integrated circuits have become fundamental in the industrial sector, driving processes in manufacturing, robotics, and smart grids. The rapid increase in the use of drones and robotics has prompted the development of more robust and reliable IC designs ~\cite{pennisi2022_ICappl_Indu}.


\textbf{\textit{Biomedical and Healthcare:}}  ICs play a crucial role in the operation of advanced medical equipment, including MRI, CT scan, and X-ray machines, making them integral to modern healthcare systems. They are also essential for the accurate and reliable performance of medical devices such as pacemakers, diagnostic equipment, and implantable sensors ~\cite{nanbakhsh2025_accelerate_longevity}. In addition, the use of wearable and implantable medical devices has expanded rapidly for various medical applications, driving significant market growth ~\cite{Eltorai2016_IntroIC_appl}.



\textbf{\textit{Computing Infrastructure:}} ICs are an essential part of computing infrastructure, driving data centers, artificial intelligence (AI) accelerators, cloud services and so on. The continual development of IC technology fosters innovation, facilitating novel applications in emerging fields such as the artificial general intelligence (AGI), edge AI, fog computing, and quantum computing, consequently shaping the future of modern computing ~\cite{golec2025_ICappl_ComputingLookingBack}.

\textbf{\textit{Aerospace and Defense:}} Integrated circuits play a key rol in navigation, radar, and secure communication systems for both sectors. Advancements in IC fabrication have facilitated more frequent and cost-effective space launches. Similar to the automotive industry, IC designs for aerospace and defense must be engineered for high durability under extreme environmental conditions, including high radiation and elevated temperatures. This has raised reliability concerns in these application domains ~\cite{santos2024_ICappli_AeroSpace, vandeburgwal2024_ICappli_aerospace}.

\begin{figure}[!t]
    \centering
    \includegraphics[width=0.95\linewidth]{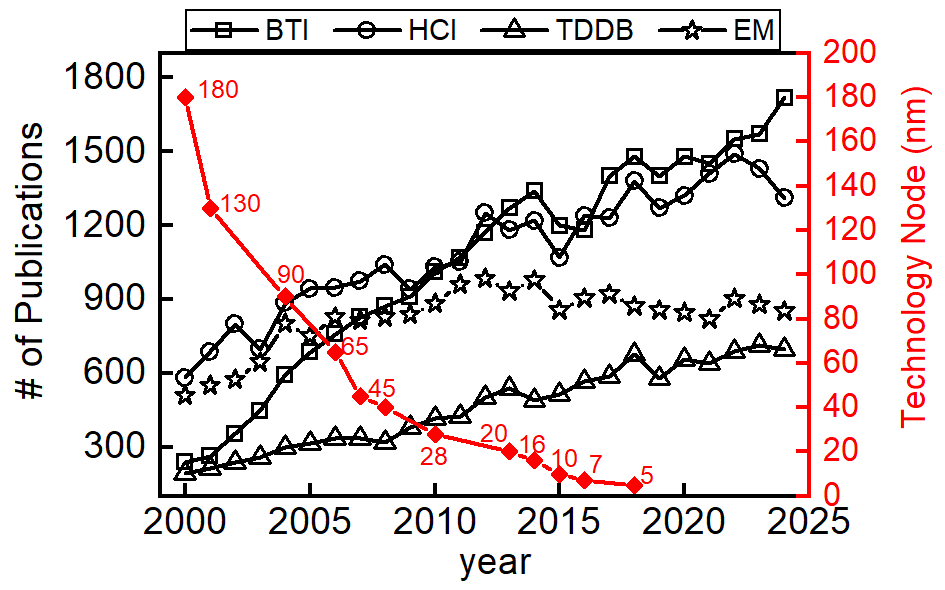}
    \caption{No. of Publications in device reliability issues and development of technology node ~\cite{demaria2021_ICIntro_technode} over past 25 years (2000-2024).}
    \label{fig:publ_vs_devRel}
\end{figure}

In summary, ICs have been widely adopted across various sectors and play a vital role in modern electronics. However, they face significant reliability challenges in nearly all of the aforementioned application domains. These challenges are exacerbated by the increasing replacement costs of ICs at advanced technology nodes and the demand for uninterrupted operation, such as during the training of AI models. Aging remains a major threat to the lifetime of IC chips, arising from a combination of device degradation mechanisms, including Bias Temperature Instability (BTI), Hot Carrier Injection (HCI), and Time-Dependent Dielectric Breakdown (TDDB), which affect transistors, as well as electromigration (EM) in on-chip metal interconnects. The confluence of these effects leads to degraded performance and shortened lifespan ~\cite{Diaz2017_CMOSRel, hill2022_CMOSRel_Intro}. The need to understand and address these aging effects has grown significantly. Figure ~\ref{fig:publ_vs_devRel} shows the number of publications addressing each aging mechanism over the past 25 years. The data were collected using the Google Scholar search engine with the following keywords: ``Bias Temperature Instability MOSFET,'' ``Hot Carrier Degradation MOSFET,'' ``Time-Dependent Dielectric Breakdown MOSFET,'' and ``Electromigration Failure Interconnects.'' The data illustrate that interest in these issues has been steadily increasing, particularly in advanced technology nodes below the 10 nm regime. This trend indicates that aging has become more severe with MOSFET scaling, highlighting a growing need to address these challenges.

\subsection{Related Surveys}
Over the last twenty years, several surveys have carefully examined aging characterization in integrated circuits, providing insights into techniques, modeling frameworks, and mitigation strategies. In the early 2000s, Intel Lab ~\cite{constantinescu2003_cirRel} analyzed key trends and challenges in circuit reliability and discussed the strategies to address them. The paper highlighted early trends and challenges in circuit reliability, emphasizing that scaling-induced variations and increased electrical fields would accelerate reliability issues. This foundational work set the stage for subsequent research into modeling and addressing aging-related failures. Various comprehensive reviews have since focused on establishing a fundamental understanding and empirical models for transistor aging mechanisms such as BTI, HCI, and TDDB. Grasser ~\cite{Grasser14_BTI_book} delivered a comprehensive review on BTI, detailing its physical mechanisms, predictive models, and circuit-level impact. Mahapatra et al. ~\cite{mahapatra2021recent} conducted a recent and detailed review on BTI and its impact on modern transistors. The study summarized the latest models, measurement techniques, and mitigation strategies, highlighting how aging impacts advanced node transistors and system performance. This work provided valuable insights into how different transistor architectures and process technologies affect BTI-induced degradation.

On the mitigation side, Rahimpour et al. ~\cite{rahimipour2012_SurveyOnChipMonitors} examined the complexity of aging-related failures in SoC design, emphasizing the importance of on-chip monitoring for real-time validation and performance adjustment. Khosavi et al. ~\cite{khoshavi2017_SurveyCMOSAging} introduced a taxonomy of aging mitigation approaches, categorizing them into worst-case design, dynamic adaptation, and adaptive resource management. Seok et al. ~\cite{seok2018_Review_Insitu_infieldMonitors} explored in-field monitoring techniques, such as voltage scaling and sensor-based feedback, to dynamically adjust system operation in response to aging-induced degradation. Halak ~\cite{Halak2020_ageICbook} examined aging mechanisms and their impact on circuit performance, presenting advanced algorithms for aging-resilient digital systems and discussing the deployment of on-chip aging monitoring sensors. Juracy et al. ~\cite{juracy2020_SurveyAgingMonitors} conducted a systematic review of aging detection and mitigation techniques, identifying timing error monitoring and voltage scaling as the most widely used approaches for real-time adjustment and mitigation of aging effects. Lanzieri et al. ~\cite{lanzieri2025_ReviewAgingTechniques} explored aging detection and mitigation strategies for Commercial-Off-the-Shelf (COTS) components, including FPGAs, microcontrollers, and SoCs. Their work highlighted the shift toward machine learning-based predictive models and real-time monitoring systems for enhanced aging resilience. Moghaddasi et al. ~\cite{moghaddasi2023dependable} reviewed reliability issues in mixed-signal circuits, focusing on aging-induced failures in analog and digital subsystems. They proposed a set of design guidelines to improve the resilience of mixed-signal SoCs. Their work also highlighted the challenges of ensuring consistent performance under aging and environmental stress, particularly in automotive and industrial applications. Recent work by Maneesha et al. ~\cite{yellepeddi2020_age_analog_AnalogCircuitDesign} investigated reliability challenges that compromise long-term performance and yield in modern semiconductor technologies. The authors discussed circuit design methodologies to counteract these issues, thereby streamlining design cycles and lowering associated costs. Afacan et al. ~\cite{afacan2021_analog_ReviewMachineLearning} and Mina et al. ~\cite{mina2022_analog_ReviewMachineLearning} reviewed advanced machine learning frameworks for automating labor-intensive stages of analog integrated circuit design. Their analyses not only identified promising avenues for future research but also presented practical case studies demonstrating how machine learning tools can address real-world design challenges from an industry practitioner’s perspective. Lienig ~\cite{lienig2018_EM_book} established foundational mechanisms and measurement techniques for evaluating EM effects in ICs. The book further emphasized the core scientific principles essential for developing a robust understanding, enabling practical application of comprehensive, system-oriented guidance for designing electromigration-aware electronic systems. 

\begin{table*}[t!]
\centering
\caption{Comparison of this survey against previous published surveys.}
\label{tab:review_comparison}
\setlength{\tabcolsep}{5pt}
\renewcommand{\arraystretch}{1.7}
\resizebox{0.85\textwidth}{!}{
\begin{tabular}{c | c c| c | c c c c | c }
\toprule
\multirow{2}*{\textbf{Ref}} & \multicolumn{2}{|c|}{\textbf{Aging Mechanism}} & \multirow{2}*{\textbf{\makecell{Aging \\Monitoring}}} & \multicolumn{4}{c|}{\textbf{Aging Characterization \& Mitigation}} & \multirow{2}*{\textbf{\makecell{Emerging  \\ Mitigation Strategies} }} \\
\cmidrule{2-3}
\cmidrule{5-8}
 & \textbf{\makecell{Device \\ Aging}} & \textbf{\makecell{Interconnect \\Aging}} &  & \textbf{Digital} & \textbf{Analog} & \textbf{SRAM} & \textbf{System} &  \\
\midrule

~\cite{Grasser14_BTI_book, Grasser15_HCI_book, mahapatra2021recent}& $\checkmark$ &  &  & $\checkmark$ &  &  & &  \\ 
~\cite{rahimipour2012_SurveyOnChipMonitors}& $\checkmark$ &  & $\checkmark$ &  &  &  & &  \\ 
~\cite{khoshavi2017_SurveyCMOSAging}& $\checkmark$ &  & $\checkmark$ & $\checkmark$ &  &  & $\checkmark$ &  \\ 
~\cite{seok2018_Review_Insitu_infieldMonitors}& $\checkmark$ &  &  &  &  &  & $\checkmark$ &  \\ 
~\cite{Halak2020_ageICbook}& $\checkmark$ &  & $\checkmark$ &  & $\checkmark$ & $\checkmark$ & &  \\ 
~\cite{juracy2020_SurveyAgingMonitors}& $\checkmark$ &  & $\checkmark$ & $\checkmark$ &  &  & $\checkmark$ &  \\ 
~\cite{lanzieri2025_ReviewAgingTechniques}& $\checkmark$ &  & $\checkmark$ & $\checkmark$ &  &  & $\checkmark$ &  \\ 
~\cite{moghaddasi2023dependable}& $\checkmark$ &  &  &  & $\checkmark$ &  & $\checkmark$&  \\ 
~\cite{mina2022_analog_ReviewMachineLearning}& $\checkmark$ &  &  &  & $\checkmark$ &  & & $\checkmark$ \\ 
~\cite{afacan2021_analog_ReviewMachineLearning}& $\checkmark$ &  &  &  & $\checkmark$ &  & &  $\checkmark$ \\ 
~\cite{yellepeddi2020_age_analog_AnalogCircuitDesign}& $\checkmark$ &  &  &  & $\checkmark$ &  & &  \\ 
~\cite{lienig2018_EM_book}&  & $\checkmark$ &  &  &  &  & $\checkmark$ &  \\ 
\makecell{This Work}& $\checkmark$ & $\checkmark$ & $\checkmark$ & $\checkmark$ & $\checkmark$ & $\checkmark$ & $\checkmark$ & $\checkmark$ \\ 
\bottomrule
\end{tabular}
}
\end{table*}

Although previous surveys have thoroughly investigated various aging mechanisms, monitoring methods, and mitigation strategies in digital systems, significant gaps remain in understanding aging across the entire IC landscape. Table \ref{tab:review_comparison} summarizes previously published reviews discussing aging mechanisms, monitoring, and mitigation strategies in ICs. One major oversight is the underexplored impact of interconnect aging, particularly electromigration, which has become increasingly severe with scaling to advanced technology nodes. Furthermore, prior surveys have largely neglected aging characterization and mitigation in analog circuits. This gap is particularly concerning given the growing reliance on mixed-signal architectures in modern integrated systems, where analog circuit degradation can significantly undermine overall system reliability. Additionally, no existing survey discussed advancements in aging research across multiple circuit domains—including digital logic, analog circuits, SRAM memory, and system-level architectures—nor do they adequately address the role of electronic design automation (EDA) techniques powered by emerging methodologies such as machine learning, graph-based learning, and approximate computing for aging analysis. This survey aims to provide a comprehensive and unified perspective on aging research, addressing these critical gaps. It revisits aging mechanisms at the device level to establish foundational knowledge for designers at all abstraction levels. It then methodically examines current monitoring frameworks and mitigation strategies at both the circuit and system levels, incorporating advanced aging characterization methods using tools such as machine learning, graph-based learning, approximate computing, and aging-aware cell libraries. Aging mitigation techniques are categorized systematically based on current computing abstractions. A particular focus is paid on aging-aware mitigation techniques developed in the last five years, which have been absent from previous reviews. This survey aims to guide future research toward developing cohesive, scalable solutions for designing aging-aware ICs and systems by bridging fragmented knowledge across different domains and emphasizing practical applications of novel techniques. The goal is to provide researchers with the foundational insights needed to address the growing need for comprehensive aging management in modern integrated circuits, ensuring sustained reliability and performance across all applications.

\section{Aging Mechanisms in Integrated Circuits}
\label{mechanisms}

\begin{figure}[!t]
\centering
\includegraphics[width=0.95\columnwidth]{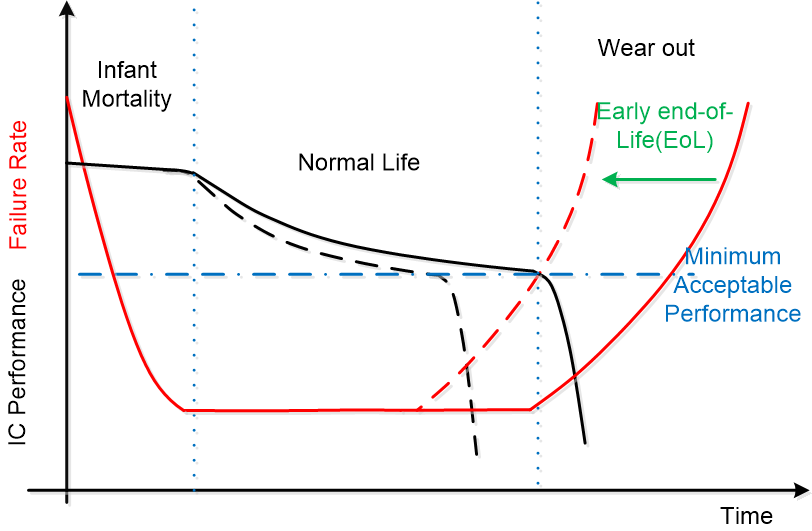}
\caption{Failure Bathtub Curve.}
\label{bathTub_curve}
\end{figure}
This section highlights the critical role of integrated circuit reliability in guaranteeing the robustness of modern electronic systems.  It analyzes advanced physics-based models for aging mechanisms, including BTI, HCI, TDDB and EM, which estimate degradation in transistors and interconnects.  These models provide accurate assessment of aging effects, including threshold voltage shifts and wire resistance changes, which degrade circuit performance. Finally, it examines how such degradations impact performance of logic circuits.  

\subsection{Overview of CMOS Integrated Circuit Reliability}
IC reliability denotes the likelihood that an integrated circuit will effectively execute its designated functions across a defined operational lifespan under specific environmental and use conditions. Hardware-level unreliability can arise from transient faults or soft errors, process variations occurs during fabrication, and aging-induced wearout failures. Soft errors, or single-event upset (SEU) caused due to and external radiation, can lead to computational inaccuracies and data corruption, however they do not impact the longevity of computer systems. Process variations are analyzed and modeled in the process design kit (PDK) to consider their effect on the circuit performance at the early design stage. However, aging-induced wearout failures might permanently degrade device parameters, necessitating early intervention during the design phase ~\cite{Maricau13_analog_rel, guo2020_ICIntro_CircadianRhythmsFuture}. Digital and analog VLSI circuits face reliability challenges stemming from both time-zero defects (e.g., process variations) and aging-induced degradation. Key device-level aging mechanisms, such as BTI, HCI and TDDB, alter critical transistor parameters (e.g., threshold voltage, leakage current), degrading circuit performance over time. Additionally, interconnect reliability issues like EM induce physical defects, such as hillocks or voids in metal lines, which can result in short circuits or open failures. The cumulative impact of these device and interconnect degradation mechanisms compromises IC functionality during operation and accelerates post-lifespan failure, underscoring the need for robust aging-aware design and mitigation strategies.

The operational lifespan of ICs is represented by the bathtub curve, which comprises three phases: infant mortality, normal life, and wear-out, as illustrated in Figure ~\ref{bathTub_curve}. During the infant mortality phase, faulty integrated circuits due to manufacturing process such as stuck-at-faults, transient faults are identified and eliminated through production testing. The normal life phase has a relatively stable failure rate; however, extended operation under rising temperatures and high supply voltages results in steady performance degradation. The wear-out/aged phase happens when performance falls below the minimum acceptable threshold, and the likelihood of failure escalates significantly after 7-10 years of operation. Recent research on degradation analysis highlights that neglecting device aging mechanisms in the design phase might expedite circuit aging, as illustrated by the dotted line in Figure ~\ref{bathTub_curve}.  Current methodologies for effective aging monitoring are insufficient; thus, integrating age-aware design strategies at the pre-silicon phase is crucial to guarantee dependable performance during the integrated circuit's intended lifespan ~\cite{yellepeddi2020_age_analog_AnalogCircuitDesign}.

\subsection{Bias Temperature Instability}
Bias Temperature Instability (BTI) is a critical reliability concern in CMOS technology, leading to gradual changes in device parameters during continuous operation. BTI degradation becomes more severe at elevated temperatures, affecting key MOSFET parameters such as threshold voltage ($V_{TH}$), subthreshold slope (SS), and transconductance ($g_m$) ~\cite{STATHIS2006_BTIReview}. The phenomenon of charge induction in SiO\textsubscript{2}, which causes BTI, was first documented in the late 1960s ~\cite{Deal1967_BTIMech}. Since then, several researchers have studied the BTI mechanism, establishing that BTI arises primarily from charge trapping at the gate oxide interface and the formation of defects within the oxide layer ~\cite{MAHAPATRA2018_BTIReview, Mukhopadhyay16_NP_BTIMech}. BTI in MOSFETs occurs in two primary forms: Negative Bias Temperature Instability (NBTI) and Positive Bias Temperature Instability (PBTI). NBTI affects pMOSFETs under negative gate bias, where holes interact with Si–H bonds at the Si/SiO\textsubscript{2} interface, creating interface traps (dangling bonds). PBTI, on the other hand, affects nMOSFETs with high-k/metal gate (HKMG) stacks and is caused by electron trapping in pre-existing oxide defects ~\cite{Mukhopadhyay16_NP_BTIMech}.


\begin{figure*}[!t]
    \centering
    \includegraphics[width=0.9\linewidth]{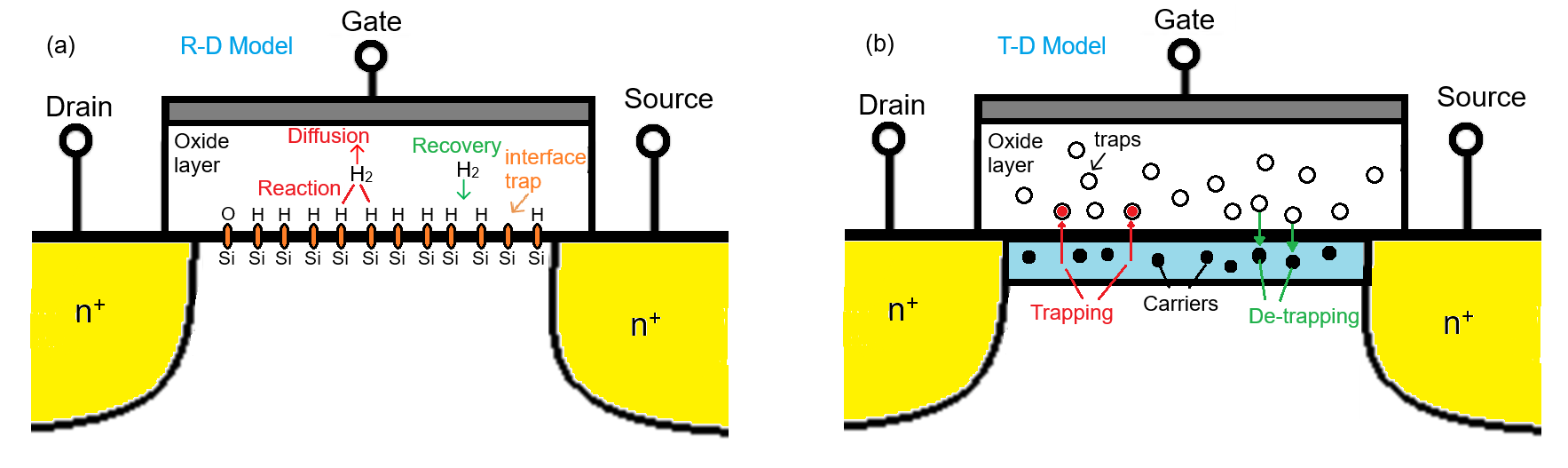}
    \caption{Illustration of BTI mechanism using (a) reaction–diffusion (R-D)  (b) trapping–detrapping (T-D) models in a p-MOSFET.}
    \label{fig:BTI_RD_TD_mech}
\end{figure*}
The BTI process has been described by two predominant models: (i) Reaction–Diffusion (R-D) and (ii) Trapping-Detrapping (T-D) ~\cite{Sutaria15_BTIMech, MAHAPATRA2018_BTIReview, guo2020_ICIntro_CircadianRhythmsFuture}. The R–D model is a bifurcated process consisting of the reaction process and the diffusion process. The R–D model suggests that stress voltage induces a breakdown of covalent bonds (Si–H) at the interface, yielding a the reaction. During the diffusion process, the dissociated hydrogen atoms combine to create H2, which then diffuses toward the gate. When gate stress is removed, the few of H2 breaks and the hydrogen atoms will recombine the dangling bonds of Si at interface, resulting in the recovery phase as shown in Figure ~\ref{fig:BTI_RD_TD_mech}(a). In modern thin-oxide devices, diffusion in the poly-gate predominates the incremental characteristics of $V_{TH}$ shift. R-D model is the foremost model used for planar MOSFETs but it fails to explains the fast recovery and AC stress effects in both nanoscale planar and finfets. Thus, an alternative trapping-detrapping (T–D) model was proposed, which indicates the presence of numerous defect states with varying energy levels and distinct capture and emission time constants. The trapping process happens when the pMOSFET is activated, altering the trap energy to capture a channel charge carrier, hence reducing the amount of accessible channel carriers, as shown in the figure Figure ~\ref{fig:BTI_RD_TD_mech}(b). The threshold voltage $V_{TH}$ increases when a trap captures a charge carrier. Upon deactivation of the pMOSFET, a passive recovery phase ensues during which some interface traps are gradually annealed, leading to partial recovery as the number of occupied traps attains a new equilibrium. This results in a partial reduction of the threshold voltage $V_{TH}$ value. The likelihood of trapping and detrapping depends on the capture and emission time constants, respectively. The decreased number of channel carriers also results in a reduction of drain current.  The AC stress in lower technology nodes have been explained with or T-D model~\cite{Reis15_CirRel_BTI_HCI}.

Among the changes in device parameters caused by BTI, the shift in $V_{TH}$ is the primary parameter identified ~\cite{MAHAPATRA2018_BTIReview}, which eventually impacts circuit performance like delay and more. Currently, an extensive unified model, an enhanced iteration of the RD model, was developed to elucidate the temporal dynamics of $\Delta V_{TH}$ under varying stress voltage, recovery voltage, frequency (f), temperature (T), and AC stress across varied pulse duty cycles (PDC) ~\cite{Desai13_BTI_Mech_DC_AC}. The model framework has been developed and validated for several advanced technologies, including Gate First (GF) HKMG FDSOI planar devices with Si and SiGe channels, Replacement Metal Gate (RMG) HKMG SOI FinFETs with Si channels, and RMG HKMG bulk FinFETs with Si and SiGe channels ~\cite{Parihar17_BTIMech_SiGe,Franco13_BTIMech_SiGe,8Parihar18_BTIMech_RMG}. Irrespective to BTI models, multiple studies indicate that the change in $V_{TH}$ due to BTI grows exponentially with continuous stress voltage and temperature, particularly in advanced technology nodes ~\cite{Reis15_CirRel_BTI_HCI,Grasser14_BTI_book,Mahapatra2013_NBTI_review}.

Multiple research teams focused on developing a compact model which can predict the degradation of transistor characteristics  over the time.~\cite{Mahapatra14_BTIMech_dev_cir,MAHAPATRA2018_BTIReview}. A compact model was therefore developed which accurately provides $\Delta$$V_{TH}$ under DC and AC stress. The developed compact model was validated, and benchmarked against various digital circuits like SRAMs and ISCAS benchmark circuits. Equation ~\ref{eq_VTHDC} shows the compact model computes the $\Delta V_{TH}$ of a MOSFET under continuous DC stress. For non-continuous AC stress, where the applied gate bias is a pulse with duty cycle $\alpha$, $\Delta V_{TH}$ is determined using Equation ~\ref{eq_VTHAC}.

\begin{equation}
    \Delta V_{TH-DC} = B *(V_G)^\Gamma * (exp(-\frac{Ea}{KT}))*t^n
    \label{eq_VTHDC}
\end{equation}

\begin{equation}
    \Delta V_{TH-AC} = \Delta V_{TH-DC} * (\alpha^{0.35}) * t^n
    \label{eq_VTHAC}
\end{equation}

Where $B$ represents technology/process dependent parameter; $V_G$ denotes the applied gate bias; $\Gamma$ represents the voltage acceleration factor; Ea indicates the activation energy (measured in eV); $k$ refers to the Boltzmann constant; $T$ represents the temperature (measured in K); $t$ is the duration of stress applied in seconds; $n$ is the time exponent for NBTI/PBTI; $\alpha$ denotes the activity factor or duty cycle of the applied gate bias ~\cite{Naphade14_BTIVari_SRAM}. $\alpha$ is regarded as 0.5 for AC stress and 1 for DC stress, respectively ~\cite{Goel15_BTIVari_SRAM}. Recent models indicate that the BTI time exponent value ($n$) is 0.17, corroborated by measurement data presented in ~\cite{Goel14_BTIMech_DCAC}. 

\begin{figure*}
    \centering
    \includegraphics[width=0.85\linewidth]{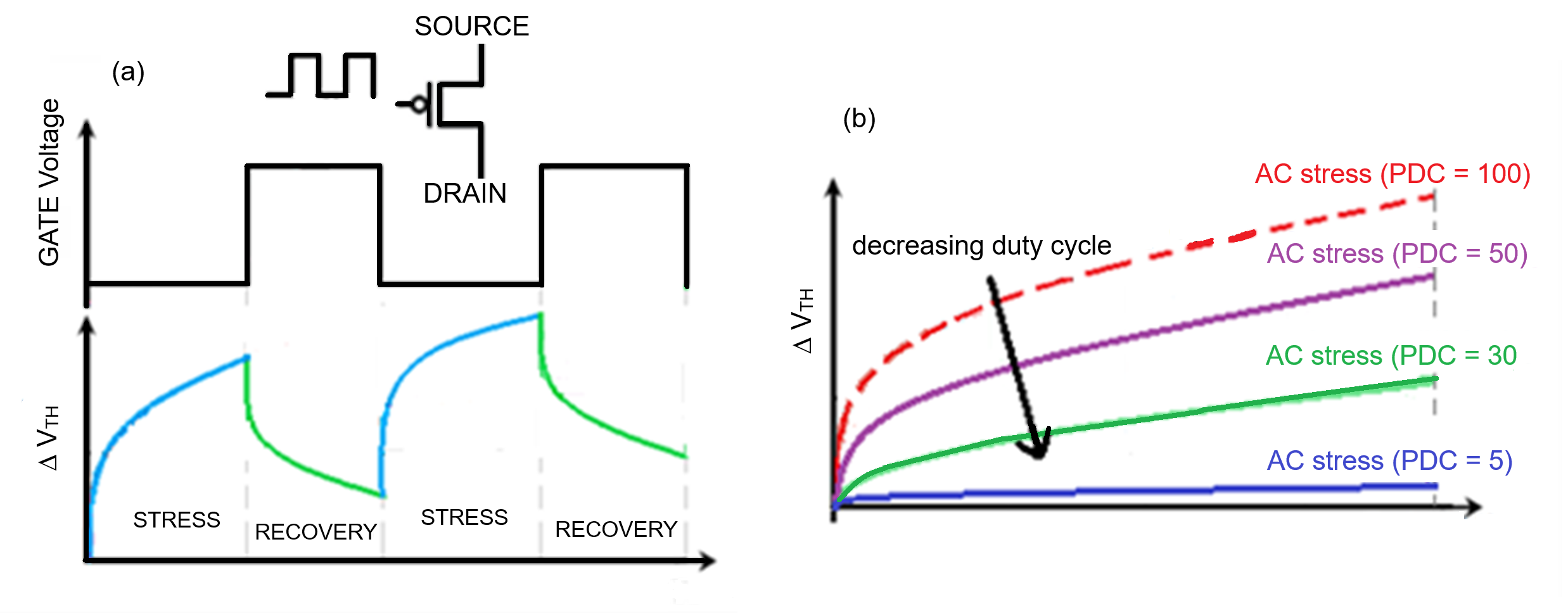}
    \caption{(a) Illustration of NBTI effect on a p-MOSFET under applied gate pulse with duty cycle (say PDC=50) (b) Shift in threshold voltage ($\Delta V_{TH}$) effect with for various PDC}
    \label{fig:BTI_PDC_stress}
\end{figure*}

Figure ~\ref{fig:BTI_PDC_stress} depicts the $\Delta V_{TH}$ in associated with the gate bias ($V_G$) applied to a p-MOSFET. For $V_G$, a pulse with a duty cycle, the MOSFET undergoes both stress and recovery phases, demonstrated by an increase in $\Delta V_{TH}$ during stress and a reduction during recovery. The figure illustrates the influence of $\Delta V_{TH}$ on the activity factor ($\alpha$) or duty cycle of applied gate bias $V_G$. When the $V_G$ is a continuous DC stress (PDC=100 or $\alpha$=1), $\Delta V_{TH}$ experiences maximal degradation. When $V_G$ is a pulse with varying duty cycles, the $\Delta V_{TH}$ corresponds to the AC stress with variable PDCs or differnet activity factor $\alpha$ values ~\cite{Chen03_BTIMech}.

\subsection{Hot Carrier Injection}
Hot Carrier Injection (HCI) poses a significant reliability issue in nanoscale MOSFETs, arising from high-energy carriers within the channel ~\cite{HUard07_NBTI_HCI_advCMOS}. The phenomenon is induced by the transversal field in the gate oxide ($E_{OX}$) and the lateral field in the channel ($E_{LAT}$), which produces high-energy (``hot'') carriers (electrons or holes) acquire the kinetic energy to surmount the $Si/SiO_2$ interface barrier adjacent to the drain region. This leads to charge trapping in the gate oxide or the development of interface defects, resulting in permanent alterations in device parameters such as threshold voltage and transconductance ($g_m$). Typically in MOSFETs with gate voltage ($V_G$) greater than half the drain voltage ($V_D$/2), the lateral electric field produces hot carriers which effects shift in device parameters ~\cite{ACOVIC1996_HCI_review,HUard07_NBTI_HCI_advCMOS}.

\begin{figure}
    \centering
    \includegraphics[width=0.95\linewidth]{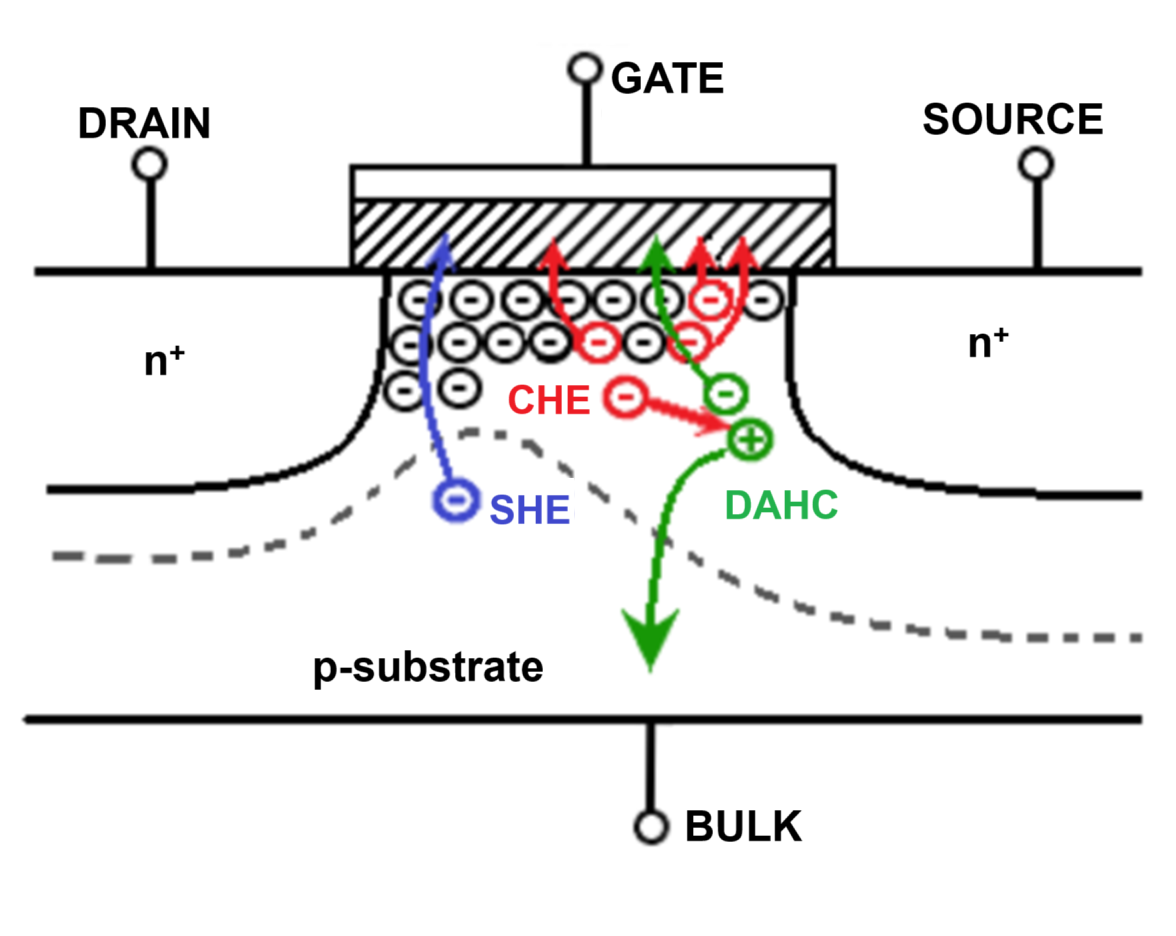}
    \caption{Illustration of different types of hot electron generation in n-MOSFET structure.}
    \label{fig:HCI_mech}
\end{figure}

Figure ~\ref{fig:HCI_mech} illustrates an n-MOSFET where hot electrons are generated based on the gate ($V_G$), body ($V_B$), and drain ($V_D$) voltages applied to the MOSFET. These hot electrons can be categorized into three types: (i) Channel Hot Electrons (CHE), (ii) Substrate Hot Electrons (SHE), and (iii) Drain Avalanche Hot Carriers (DAHC) ~\cite{Maricau13_analog_rel}. Hot electrons can be produced in the gate oxide, leading to the formation of oxide traps and interface traps, as shown in Figure ~\ref{fig:HCI_mech}. These traps alter key device characteristics, including threshold voltage ($V_{TH}$), linear drain current ($I_{DLIN}$), and saturation drain current ($I_{DSAT}$). Among the three mechanisms, CHE and DAHC have a more pronounced impact on transistor degradation compared to SHE, thereby limiting the device’s lifespan and affecting circuit performance. A similar mechanism applies to p-MOSFETs, where holes serve as the channel carriers ~\cite{PARTHASARATHY06_devRel_advCMOS}. Channel hot carriers influence carrier mobility; since holes are significantly heavier than electrons, they experience fewer collisions during avalanche multiplication, resulting in reduced hole injection into the gate oxide. Consequently, HCI degradation is more prevalent in n-MOSFETs than in p-MOSFETs, as reported in the literature ~\cite{PARTHASARATHY06_devRel_advCMOS}.


The impact of HCI on CMOS circuits has been studied for over four decades. In 1979, Hu et al. proposed the Lucky-Electron Model (LEM), which provides a closed-form expression for gate current and substrate current resulting from the hot carrier effect ~\cite{Hu1979_HCI_model}. Takeda et al. ~\cite{Takeda1983_HCI_model} established an empirical correlation among the change in threshold voltage ($\Delta V_{TH}$), substrate current ($I_{SUB}$), and drain voltage ($V_D$) based on experimental data. Since then, numerous HCI models based on LEM have been introduced in the literature, each exhibiting distinct analytical formulations due to advancements in MOSFET technology ~\cite{ACOVIC1996_HCI_review}. However, these LEM-based models primarily account for channel hot carriers while neglecting other HCI mechanisms that also contribute to device parameter degradation. Consequently, alternative advanced HCI models have been proposed ~\cite{PARTHASARATHY06_devRel_advCMOS}. Wang et al. formulated an HCI compact model derived from LEM, but it was limited to the reliability simulation of digital circuits at the nanoscale ~\cite{Wang22_HCIMech_Simu}. Subsequently, Maricau et al. developed an HCI model addressing both static and dynamic channel hot carrier (CHC) effects on device parameters ~\cite{Maricau13_analog_rel}, deriving an empirical equation based on the LEM model. The deterioration in threshold voltage is represented by a power-law dependence on the duration of applied stress ~\cite{MARICAU2008_HCI_model}, as expressed in following Equation.

\begin{multline*}
\Delta V_{TH-HCI} = C_{HCI}  \left [ \left (\frac{1}{L}\right)^n \cdot (V_{GS}-V_{TH0})^{1-n} \right]\cdot \\ \left [exp(\alpha_{1} E_{OX}) \cdot  exp\left (-\frac{\alpha_2}{E_{LAT}}\right) \cdot  exp\left (-\frac{E_a}{KT}\right)\right]t_{str}^n
    \label{eq_HCI_VTH}
\end{multline*}

where $C_{HCI}$, $\alpha_1$, and $\alpha_2$ are technology-dependent parameters; $E_a$ denotes the activation energy (in eV); $K$ represents the Boltzmann constant; $T$ signifies temperature (in K); $E_{OX}$ and $E_{LAT}$ refer to the oxide and lateral electric fields, respectively; $V_{GS}$ and $V_{TH0}$ indicate the applied gate voltage and the time-zero threshold voltage; $L$ denotes the channel length; $t_{str}$ is the applied stress time; and $n$ is the time exponent in the HCI model, which varies across technologies, typically ranging from 0.3 to 0.7 ~\cite{MARICAU2008_HCI_model, PARTHASARATHY06_devRel_advCMOS, Wang07_CirRel_model}. Maricau et al. validated their HCI model using experimental data from 65nm technology ~\cite{MARICAU2008_HCI_model}. More recently, Mahapatra et al. examined HCI degradation under various experimental settings, including combined DC-AC stress, using improved MOSFET technology. Their research group developed a unified compact model to predict the degradation of device properties resulting from HCI ~\cite{Mahapatra20_HCIreview_p1, Mahapatra20_HCIreview_p2, Sharma19_HCI_model}. Currently, only a few research groups are working on a unified model that integrates the effects of both HCI and BTI into a single framework ~\cite{Ullmann19_Comp_HCINBTI_model_p1, Jech19_Comp_HCINBTI_model_p2}. 

\subsection{Time Dependent Dielectric Breakdown}
\begin{figure}
    \centering
    \includegraphics[width=0.95\linewidth]{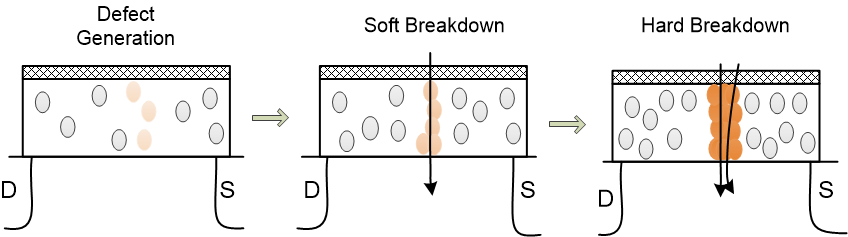}
    \caption{Cross-section of a transistor with the gate oxide traversing all three stages of dielectric breakdown.}
    \label{fig:TDDB_mech}
\end{figure}

Time-Dependent Dielectric Breakdown (TDDB) is a critical reliability failure mechanism in MOSFETs, characterized by continuous electric field stress across the gate oxide, leading to the formation of a conductive channel through the dielectric, popularly referred as dielectric breakdown ~\cite{McPherson1985_TDDBMech,Lloyd05_TDDBMech_model,Suhele02_TDDBMech_model}. It is the gradual and eventual breakdown of a gate oxide dielectric ( also known as gate oxide). The fundamental physical mechanism underlying TDDB entails the formation of defects and the entrapment of charge in the dielectric. Under continuous voltage stress, energetic charge carriers (electrons or holes) tunnel through the dielectric by Fowler–Nordheim or trap-assisted mechanisms, interacting with the atomic bonds of the dielectric ~\cite{McPherson12_TDDB_review}. The TDDB degradation mechanism transpires in three distinct phases ~\cite{Nigam13_Moore_Rel}, as illustrated in Figure ~\ref{fig:TDDB_mech}: (i) Defect Generation: Under strong electric fields (5–10 MV/cm), covalent bond breakdown occurs, leading to the formation of traps (e.g., oxygen vacancies, dangling bonds) inside the oxide layer. The trapped charges in the dielectric and at the SI/SIO2 interface subsequently degrade the device characteristics, notably the threshold voltage $V_{TH}$. (ii) Soft Breakdown: The gradual accumulation of defects leads to a conductive percolation path connecting the gate and substrate. This creates a continuous conduction channel that increases leakage current and reduces the device's switching speed. (iii) Hard Breakdown: The elevated temperature due to the leakage current results in the formation of additional traps, creating a wider and less resistant path. This leads to the increased gate current, resulting in thermal runaway that entirely breaks the dielectric layer.

The primary factors accelerating TDDB in MOSFETs are the applied electric field, temperature, and defects in the material. The applied electric field is the primary factor in defect or trap generation, which is further amplified by thermal energy. With technological advancements, the fabrication of thinner gate oxides and high-K dielectrics (e.g., HfO\textsubscript{2}) has led to increased defect densities ~\cite{Suhele02_TDDBMech_model,Degraeve1995_TDDBMech_model}. The literature presents TDDB models based on either field-induced degradation, current-induced degradation, or a combination of both mechanisms ~\cite{McPherson12_TDDB_review}. Researchers have also examined various assumptions on defect-generation kinetics and conduction mechanisms for developing these models ~\cite{Lloyd05_TDDBMech_model}. The four major models commonly used to characterize the time to failure (TTF) using the TDDB mechanism are as follows:
\begin{itemize}
    \item \textbf{Thermochemical (E-) Model:} This model attributes degradation to the electric field, suggesting that bond breakdown is accelerated by the oxide electric field ($E_{ox}$) and temperature ($T$). It primarily analyzes the time to failure in thick-oxide (SiO$_2$) dielectrics.  
    \item \textbf{Anode Hole Injection (1/E) Model:} This current-based degradation model indicates that the breakdown time decreases exponentially with the inverse of the electric field. It is often applied to high-K dielectrics.  
    \item \textbf{Power-law Voltage ($V^N$) Model:} Also known as the anode hydrogen release (AHR) model, it posits that the Si-H bond near the Si/SiO$_2$ interface is stimulated by single or multiple electron events. It is frequently used for ultrathin dielectric MOSFETs (under 4 nm high-K dielectrics).  
    \item \textbf{Exponential ($E^{1/2}$) Model:} This current-induced model states that degradation results from current flow through the dielectric. It is commonly employed for low-K SiO$_2$-based interconnect dielectrics.  
\end{itemize}

McPherson ~\cite{McPherson12_TDDB_review} have explained each TDDB mechanism and models in details. Aging due to TDDB causes the  eventual failure of dielectric structure of the device, leading to permanent failure of the switching operation of the device in a circuit. Thus, it is important for designers to carefully model the time to failure (TTF) from one of the above mentioned models.     

\begin{figure*}[!t]
    \centering
    \includegraphics[width=0.75\linewidth]{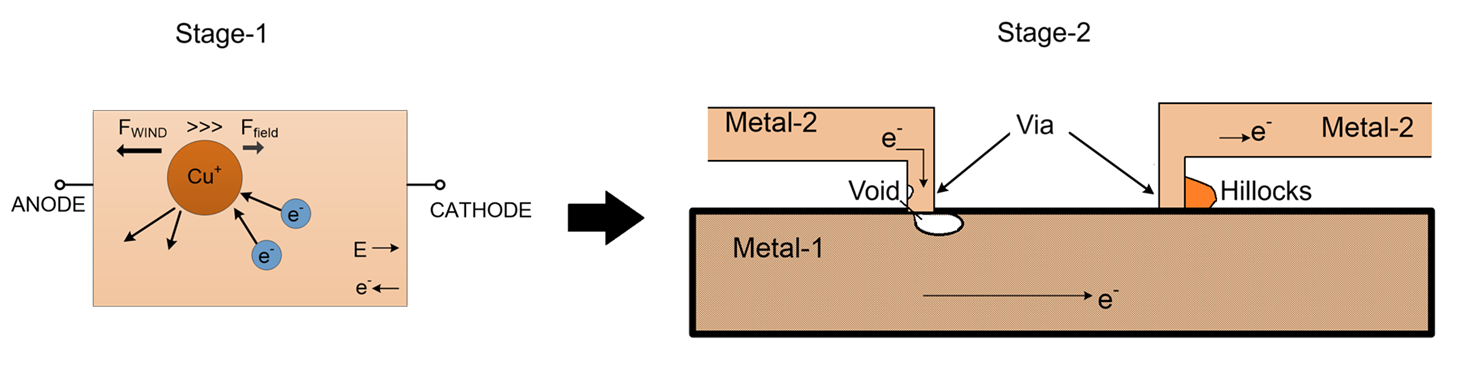}
    \caption{Illustration of forces acting on metal ions (Cu+), where the momentum transfer from the electrons. Formation of voids and hillocks near via connection between two metal lines resulting a failure mechanisms due to EM in integrated circuits.}
    \label{fig:EM_mech}
\end{figure*}

 \subsection{Electromigration (EM)}

Electromigration (EM) is a significant failure mechanism in integrated circuit interconnects, caused by the displacement of metal atoms under elevated current densities. This atomic displacement leads to a formation of voids (vacancies) that induce open-circuit failures or hillocks (metal accumulations), which cause short circuits. These structural defects ultimately compromise interconnect reliability, leading to circuit failures ~\cite{Hu18_EM_mech,lienig2018_EM_book,Petrescu04_EM_Model}. 

Figure ~\ref{fig:EM_mech} depicts the electromigration (EM) failure mechanism, which occurs in two stages.  The first stage involves momentum transfer, wherein electron wind transfers kinetic energy to metal ions (e.g., Cu or Al), pushing them in the direction of current flow.  The second is the divergence stage, in which atomic flux disperses across grain boundaries, vias, or interfaces, resulting in the creation of voids or hillocks. These structural anomalies—voids at grain boundaries and hillocks between adjacent metal lines—result in increased interconnect resistance and short circuits, respectively ~\cite{young1994_EM_FailureMechanism, zhao2022_EM_review_RecentProgress, marti2016_EM_MultiphysicsModel}. 

The primary factors governing EM in integrated circuits are current density (J) and operating temperature (T) . Black’s empirical equation ~\cite{Black1969_EM_model}, widely used to predict interconnect reliability, quantifies the mean time to failure (MTTF) as given in equation (~\ref{EM_MTTF}). 

\begin{equation}
   MTTF = \frac{A}{J^n} exp \left ( \frac{E_a}{KT} \right)
   \label{EM_MTTF}
\end{equation}
where A is a constant depends on material properties and interconnect cross-sectional area; J is the current density; n refers to current density exponent (~1–2 for Cu, ~2–3 for Al),
$E_a$ is the Activation energy for atom diffusion (~0.8–1.2 eV for Cu). Blech ~\cite{Blech1976_EM_model} presents a critical current density, below which EM-induced failure is mitigated. This blech limit applies to short interconnect segments and is less predictable for complicated interconnect trees. EM is particularly crucial for modern integrated circuits, where the reduced dimensions of the interconnect and the higher current densities worsen the problem.

\subsection{Stochastic Aging-induced Variations}

MOSFET devices undergo degradation mechanisms, including BTI and HCI, which result in shifts in the threshold voltage ($V_{TH}$) over prolonged usage. These changes are naturally stochastic due to their origin in microscopic processes (e.g., trap formation and charge capture) that differ from one device to another. The resultant stochastic variations in $V_{TH}$ throughout a chip can impair performance, yield, and long-term reliability ~\cite{Rauch02_BTIVari,Rosa06_BTIVari}. Under prolonged bias stress (usually negative for pMOS devices, referred to as NBTI), silicon-oxide interfaces accumulate a stochastic quantity of interface traps. These traps are formed by  breaking of Si-H bonds followed by reaction-diffusion mechanisms. The location and density of these traps fluctuate stochastically, resulting in a non-uniform threshold voltage shift ($\Delta V_{TH}$) across devices ~\cite{Rauch07_BTIVari}. Similarly, high-energy carriers formed at the drain region can break Si–H bonds at the interface, resulting in interface states that can vary $V_{TH}$. Since the occurrence of such ``lucky'' events follows to a probabilistic structure, HCI also induces variability in threshold voltage shifts over time ~\cite{PROCEL2015_HCIVari}. The underlying mechanisms (trap creation, defect reactions, and recovery dynamics) for both NBTI and HCI are stochastic at the atomic scale, hence the resultant $V_{TH}$ shifts ($\Delta V_{TH}$) are most accurately characterized by statistical distributions. This statistical behavior is crucial for designers, as it directly influences variability and discrepancies in large-scale circuits ~\cite{Angot13_BTIVari_EP, Goel15_BTIVari_SRAM, Naphade14_BTIVari_SRAM}.

 Negative BTI (NBTI) is studied to be more prone to device parameter degradation than positive BTI (PBTI) at lower technology nodes. NBTI-induced variability, also known as NBTI variability, is recently getting more attention at advanced CMOS technology nodes ~\cite{Lee17_BTIVari_DS, Kerber13_BTIVari_DS, MUKHOPADHYAY2018_BTIVari_DS, Kaczer10_BTIVari,Angot13_BTIVari_EP, Prasad16_BTIVari_Gamma, Liu15_BTIVari_Norm}. Limited study has been conducted on HCI-induced variability, HCI variability ~\cite{Makarov20_HCIVari, Ma2014_HCIVari, Bottini18_HCIVari}. There are several reports presented on effect of NBTI variability circuit performance. Therefore, circuit designs should make resilient to NBTI variability. In the past decade, various statistical models have been proposed to model BTI induced variations. Statistical compact models present the NBTI induced variations as $\Delta V_{TH}$ variations, and it is crucial to define appropriate design margins for guaranteed circuit performance until End-of-Life. Rauch et al. from IBM characterized the NBTI-induced variations as $\Delta V_{TH}$ variations utilizing the Dispersive Skellam (DS) distribution ~\cite{Rauch07_BTIVari, Lee17_BTIVari_DS, Kerber13_BTIVari_DS, MUKHOPADHYAY2018_BTIVari_DS}. Kaczer et al., from IMEC, presented an analytical model of $\Delta V_{TH}$ variability characterized by an exponential Poisson (EP) distribution ~\cite{Kaczer10_BTIVari, Angot13_BTIVari_EP, Prasd14_BTIVari_EP}. This anlytical model was then updated at lower technology nodes by using the Gamma distribution ~\cite{Prasad16_BTIVari_Gamma, Naphade14_BTIVari_Gamma, Goel15_BTIVari_SRAM}. Liu et al., from Samsung Electronics Co. Ltd., analyzed the experimental data of FinFET technology and modeled $\Delta V_{TH}$ variability using Normal distributions ~\cite{Liu15_BTIVari_Norm, Liu20_BTIVari_Norm}.
\begin{figure*}[!t]
    \centering
    \includegraphics[width=0.7\linewidth]{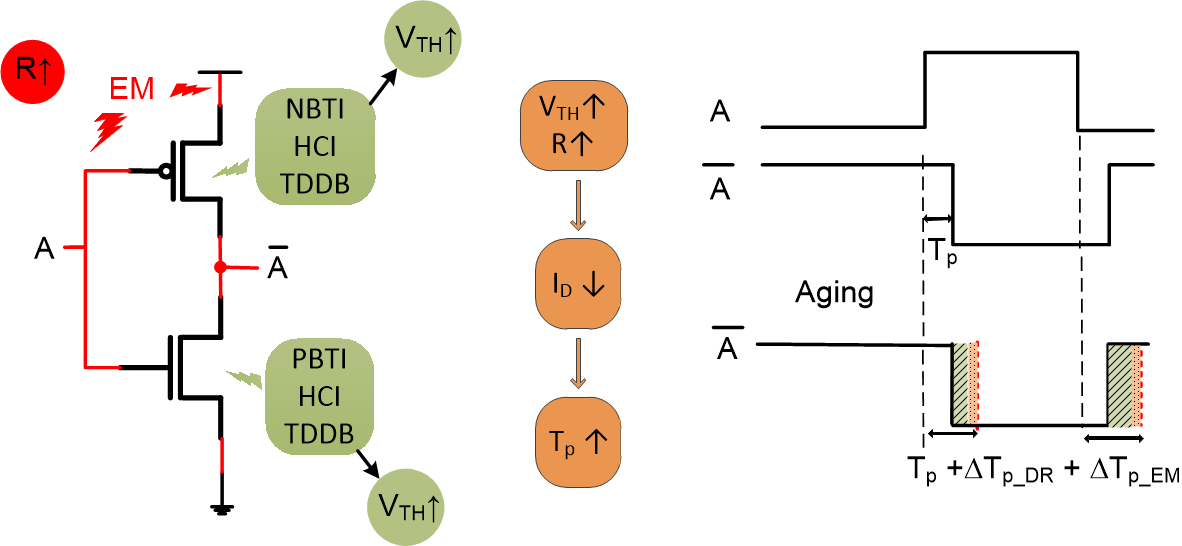}
    \caption{ Illustration of device and interconnect reliability issues in a CMOS inverter circuit. Device reliability (DR) shifts the threshold voltage of a transistor, and EM increases the resistance  of a metal wire. Also demonstrates the shift in device parameter effecting the propagation delay ($T_p$).}
    \label{fig:effect_cirPerfo}
\end{figure*}
\subsection{Effect of Aging Mechanisms at Circuit level}
Device reliability (DR) issues, including the aforementioned BTI, HCI, TDDB, and EM effects, significantly degrade circuit performance. These DR issues primarily degrade transistor performance by shifting the threshold voltage, which reduces the drive current ($I_D$) in transistors, directly impacting circuit timing. Additionally, EM in interconnects increases the metal resistance over time, further degrading performance. 

Figure ~\ref{fig:effect_cirPerfo} illustrates how combined device-level and interconnect-level reliability issues impact the propagation delay ($T_p$) of an inverter circuit. As BTI and HCI increase the threshold voltage ($V_{TH}$), the discharge current through the MOSFETs decreases, slowing the transitions from high-to-low and low-to-high. Simultaneously, the increase in resistance due to EM in the power delivery network (PDN) or signal paths leads to an IR drop, effectively lowering the supply voltage ($V_{dd}$) and further reducing the drive current. The overall propagation delay of an aged inverter circuit can be modeled by the additive effects of device-level and interconnect-level degradations as  
\begin{equation}  
T_{p,aged} = T_{p,time0} + \Delta T_{p\_DR} + \Delta T_{p\_EM} 
\end{equation}  
where $T_{p,aged}$ is the propagation delay after aging, $T_{p,time0}$ is the initial delay, $\Delta T_{p\_DR}$ is the delay increase due to device reliability issues, and $\Delta T_{p\_EM}$ is the additional delay from electromigration. Waveforms shown in Figure ~\ref{fig:effect_cirPerfo} demonstrate that the combined aging effects progressively worsen inverter performance over time.   

Currently, semiconductor foundries such as TSMC provide compact models to examine device degradation and its impact on circuit performance ~\cite{Lee2014_TMI_TSMC}. EDA companies such as Cadence provides industry-standard reliability simulation tools (Cadence RelXpert) and a reliability model file (AgeMOS model) for assessing the effects of HCI on both analog and digital circuits ~\cite{Maricau13_analog_rel, art_AnalogRelSim_ageMOS}. Similarly, Synopsys provides the reliability simulation tool with model file MOSRA for aging characterization in CMOS circuits ~\cite{Tudor2010_relSim_MOSRA}. The degradation in circuit performance caused by aging mechanisms highlights the essential requirement for the implementation of aging monitoring circuits and techniques at the system level. These monitoring techniques detect degraded circuits and provide critical insights for aging-aware approaches to design. The next section discusses recent improvements that highlight the significance of monitoring circuits, as they enhance the reliability of integrated circuits by addressing the growing reliability issues in modern semiconductor technologies.

\section{Aging Monitoring Circuits and Systems}
\label{monitor}
As discussed in the last section, aging-induced degradation affects critical transistor characteristics (e.g. threshold voltage $V_{TH}$) or interconnect resistance, resulting in a longer propagation delay and ultimately affecting the performance of ICs. Aging monitoring circuits are crucial for detecting and analyzing the impacts of reliability issues, enabling designers to take steps to mitigate potential issues before they lead to system-level failures. 

Multiple review papers have addressed the aging monitoring techniques and sensor circuits. ~\cite{rahimipour2012_SurveyOnChipMonitors} reviewed on-chip monitors focusing on soft-errors, temperature, critical path, and aging monitors. ~\cite{khoshavi2017_SurveyCMOSAging} reviewed aging monitors, aging models, and aging mitigation techniques, primarily focusing on digital circuits. ~\cite{juracy2020_SurveyAgingMonitors} analyzed different aging monitors and reconfiguration techniques. Aging monitoring techniques have been explored in the literature with a primary focus on digital circuits and systems. This survey primarily focuses on the monitoring techniques associated with the fundamental methodology used to identify the degraded device or circuit performance. Thus, this section presents techniques used to identify degradation in device properties (such as $V_{TH}$, $I_{D}$ and $R$) and circuit performance metrics, including path delay. Consequently, we survey ring oscillators, critical path replicas, aging monitoring sensor-based circuits and EM sensing schemes.

\renewcommand{\arraystretch}{1.5}
\begin{table*}[t]
\centering
\caption{Literature on Ring-Oscillator (RO), Critical Path Replicas, Analog aging monitor sensor Circuits fro aging monitoring in ICs.}
\label{Table:agingMoni_liter}
\resizebox{0.65\textwidth}{!}{
\begin{tabular}{c|c}

\toprule
\textbf{Monitoring Technique} & \textbf{Previous Published Works} \\

\midrule

\textbf{Ring-Oscillator (RO)} & ~\cite{park2021_ageMoni_RO,Gondo2020_ageMoni_RO,Igarashi15_ageMoni_RO, Morsali2022_ageMoni_RO, Sengupta2017_ageMoni_RO}\\

\textbf{\makecell{Critical Path Replicas (CPR)}} & ~\cite{Gomez2016_ageMoni_CPR, Ashraf2015_ageMoni_CPR, Lorenz2014_ageMoni_CPR, wang2024_ageMoni_InSituCriticalPathReplica, changho2021_ageMoni_CPR_variation, guo2015mcpens}\\

\textbf{\makecell{Monitoring-based Aging Sensor Circuits}} & ~\cite{Karimi2017_ageMoni_sens, Moustakas_ageMoni_VTHSensor, Lin2018_ageMoni_sens, Bhootda22_ageMoni_sens_VTH, zhang2022_ageMoni_sens, Wang2019_ageMoni_sensor, Rohbani2018_ageMoni_sens, Vargas24_ageMoni_sens_finFet, Shah2017_ageMoni_BTISensor}\\

\textbf{EM Sensor} & \cite{Wang2011_ageMoni_EMsens, Slottke2015_ageMoni_EMsens_wafer, guo2020_ICIntro_CircadianRhythmsFuture, he2015based}\\

\bottomrule
\end{tabular}
}
\end{table*}

\subsection{Ring Oscillators (RO)- based Monitors}
A ring oscillator (RO) is built with an odd number of inverters configured in a closed loop that ensures oscillation. The frequency of oscillation $f_{osc}$ can be found in Equation ~\ref{eq_fre_osc_RO}, where $N$ is number of inverter stages, and $t_p$ is the single-stage delay.
\begin{equation}
    f_{osc} = \frac{1}{2N\cdot t_p}
    \label{eq_fre_osc_RO}
\end{equation}
As aging mechanisms such as BTI and HCI accumulate, the oscillator's frequency reduces. Experimental data indicates that accelerated life testing, induces measurable frequency shifts that correlate with established aging parameters. With advanced BTI and HCI models, we can correlate the shift in frequency due to degradation in MOSFET parameters ~\cite{lanzieri2025_ReviewAgingTechniques}. These RO-based aging monitors are compact and can be inserted in several locations with minimal design disruption. They also offer real-time monitoring, enabling early detection and subsequent adaptation. The technique is frequently used in integrated circuits due to its low complexity and capability for real-time monitoring. Several studies have been explored in the RO-based aging monitoring technique. An on-die aging monitor using ring oscillator to assess BTI and AC-induced HCI was proposed ~\cite{Igarashi15_ageMoni_RO}. The monitor comprises a symmetric RO and an asymmetric RO, emphasizing sensitivity to DC stress and speed degradation induced by AC-HCI. The monitor enhances design guard bands, facilitating reliable components in high-performance application ICs. In ~\cite {tsiatouhas2017_ageMiti_PeriodicBTI}, the frequency of a local ring oscillator is exploited for BTI-induced aging monitoring in SRAM memory cells. However, this frequency is also prone to process and temperature variations. ~\cite{park2021_ageMoni_RO} proposed an on-chip reliability monitor evaluates all four sources of BTI through the use of strained ring oscillators and beat frequency detecting methods. It attains a frequency measurement resolution of 0.01\% and 4 microseconds. ~\cite{Sengupta2017_ageMoni_RO} introduced an innovative technique utilizing on-chip sensors based on ring oscillators to identify delay shifts caused by aging in nanometer-scale circuitry. The method employs presilicon analysis to calculate calibration factors, enabling precise delay estimations within 1\% of actual values.  A refinement mechanism for post-silicon analysis partially captures the circuit's load, yielding an 8\% reduction in delay guardbanding overheads relative to traditional methods. Similarly, ~\cite{Gondo2020_ageMoni_RO} introduced an aging-resistant RO as a digital temperature and voltage sensor, showcasing its efficacy in alleviating delay degradation and minimizing sensor measurement inaccuracies. In summary, the degraded value frequency of oscillation for RO can be used to identify the aging-related issues in ICs. However, RO-based aging sensors have a few limitations, including the possibility for variations in ambient temperature and supply voltage to distort measurements unless environmental factors are taken into consideration. Additionally, periodic recalibration is necessary to confirm that frequency shifts are attributed solely to aging. This has led to development of other types of aging monitors.

\subsection{Critical Path Delay Monitors}
Critical Path Replica (CPR) is a widely used circuit-based aging monitoring technique that emphasizes the replication of the critical paths of a circuit under test (CUT) to identify and measure aging-related performance degradation. The critical path of a digital circuit denotes the longest combinational logic sequence between registers, which governs the circuit's maximum operating frequency. These paths are replicated via specialized delay chains that mimic the original circuitry. A comparator or latch system is employed to assess the delay of the replicated path with respect to a reference path delay value. Any increase in delay over a predetermined threshold signifies age-related degradation. The delay $t_{delay}$ can be represented as $t_{delay} = t_0 + \Delta t_{age}$, where $t_0$ represents the nominal time and $\Delta t_{age}$ denotes the additional delay due to aging. BTI aging models enable designers to correlate $\Delta t_{age}$ with operational stress and time ~\cite{juracy2020_SurveyAgingMonitors}. 

Multiple studies demonstrate the use of critical path replica for aging monitoring and variation-aware designs in integrated circuits ~\cite{changho2021_ageMoni_CPR_variation, wang2024_ageMoni_InSituCriticalPathReplica}. ~\cite{Lorenz2014_ageMoni_CPR} suggested a method for better-than-worst-case design that includes periodic monitoring and circuit degradation countermeasures. Their primary contribution is an algorithm designed to identify possible critical paths (PCPs) over the designated lifespan, taking into account local process variations. This method decreases the monitored paths by a factor of 2.7 relative to existing techniques, guaranteeing the identification of all possible critical paths. The Reactive Rejuvenation (RR) architectural technique aims to minimize the aging of ICs caused by BTI by detecting and adjusting timing violations ~\cite{Ashraf2015_ageMoni_CPR}. This lightweight logic circuit continuously assesses the BTI effect on the performance of critical and near-critical paths, restoring the timing-sensitive segment of the circuit via switching computations to a redundant aging-critical voltage domain. The technique accomplishes the mitigation of aging and the reduction of energy use. In ~\cite{guo2015mcpens}, an aging sensor based on a metastable cell was proposed. It comprises a degradation inverter and a reference inverter. Driven by the paths of interest, the sensor is exposed to the same environment as the paths and degrades at the same rate as these paths. A detailed review of critical path-based aging monitor circuits has been presented in ~\cite{juracy2020_SurveyAgingMonitors}. To summarize this technique, the degraded critical path value is compared with the non-degraded replica circuit value to estimate circuit degradation.

\subsection{Monitoring-based Aging Sensor Circuits}
As aging-related issues degrade device characteristics, aging sensors have been developed to measure MOSFET degradation (e.g., $V_{TH}$ and $I_{ON}$ reduction) directly. They provide quantifiable information on physical degradation mechanisms such as BTI, HCI, and TDDB, enabling early-warning functionality for adaptive system responses. Though monitoring-based aging sensors are effective, challenges of this type of sensors include sensitivity to process variations and the need for effective compensation ~\cite{Ren2025_ageMoni_sens_RealTime}. Efforts have been paid to enhance accuracy and robustness to ensure reliable operation across different environments. 

Several novel circuits have been proposed to characterize BTI-induced degradation ~\cite{Shah2017_ageMoni_BTISensor, shah2018_ageMiti_resiliCir, Shah2018_ageMiti_SRAM, Bhootda22_ageMoni_sens_VTH, Gupta2021_ageMiti_digi}. A real-time CMOS $V_{TH}$ sensor using a nested feedback loop in 0.18 $\mu$m CMOS technology demonstrates strong supply voltage tolerance and self-compensation for second-order effects, achieving a maximum $V_{TH}$ deviation error of 0.25\% ~\cite{Moustakas_ageMoni_VTHSensor}. Online aging monitoring is essential for Insulated Gate Bipolar Transistors (IGBTs). A leakage current ($I_{leak}$) sensor measures $I_{leak}$ online with high sensitivity, simplicity, and cost-effectiveness ~\cite{zhang2022_ageMoni_sens}. A write current-based BTI sensor (WCBS) evaluates SRAM BTI aging by tracking maximum write current variations, with ±1.25 mV accuracy (±3.2\% error) ~\cite{Karimi2017_ageMoni_sens}. BTI evaluation precision is improved by programming specific bit patterns. 

\subsection{EM Sensing}
Compared to transistor aging sensing, EM sensing is more challenging for several reasons. First, EM behavior is more complex; the resistance remains stable over a long stress accumulation period, followed by a sudden increase. This necessitates more frequent EM sensor activation than BTI sensors to capture EM-triggered events accurately and promptly. Second, EM primarily occurs in the power delivery network (PDN), which is distant from the logic in the front-end-of-line, making it difficult for logic-based sensors to replicate the same environmental conditions as top-layer metal interconnects. Finally, distinguishing EM-induced degradation from other aging mechanisms like BTI is difficult since both cause similar performance degradation (e.g., slowdown).

\begin{figure}[!t]
\centering
\includegraphics[width=0.45\textwidth]{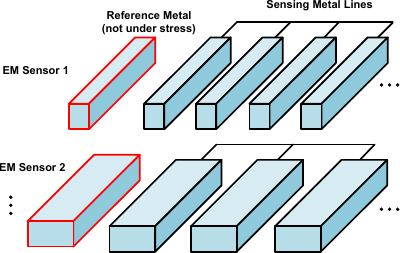}
\caption{Illustration of Metal-line-based EM sensors. Multiple dimensions can be used to sense at different levels ~\cite{guo2020_ICIntro_CircadianRhythmsFuture}.}
\label{EM_Sensing}
\end{figure}

EM monitoring techniques typically use conductive traces along integrated circuits. A current is passed through these traces, and the shift in resistance is measured by monitoring the IR drop ~\cite{allee2004_ageMoni_Em_dete_patent, chen2015diagnostic, he2015based}. The main idea is to place on-chip metal lines that experience comparable or greater EM stress than the PDN or other EM-sensitive interconnects. Figure~~\ref{EM_Sensing} illustrates an example of such an EM sensing structure, based on previous work ~\cite{he2015based}. It consists of parallel on-chip metal lines, with a reference metal line held unstressed to detect resistance changes. Multiple metal lines are required to account for inherent variations and obtain statistically meaningful results. Two standard wafer-level EM monitoring techniques have been proposed: SWEAT (Standard Wafer-level Electromigration Acceleration Test) and BEM (Breakdown Energy of Metal) ~\cite{Fujii1995_ageMoni_EM_sens_SWEAT_BEM}. However, these methods are limited to detecting EM-induced voids. A novel on-chip current measurement technique for EM management utilizes the voltage drop across existing interconnects to measure current flow. A MOSFET-only sensing circuit provides 9-bit resolution at midrange current levels, offering resilience to local variations and suitability for multi-site on-chip current assessments ~\cite{Wang2011_ageMoni_EMsens}.

\subsection{Summary of Aging Sensing Techniques}
The literature on aging monitoring techniques is summarized in Table ~\ref{Table:agingMoni_liter}, which organizes key references according to the specific monitoring techniques discussed above. Aging sensor circuits may experience self-degradation over time, leading to inaccurate measurements and false detections. This degradation can result from various factors, including BTI or EM, similar to the aging effects they are designed to detect. Future research should focus on developing integrated monitoring systems that combine aging sensors with conventional monitoring methodologies. By integrating multiple approaches, more robust and accurate monitoring solutions can be developed, addressing the limitations of single techniques and improving the long-term reliability of ICs. Given the significant challenges posed by aging effects, exploring mitigation strategies to reduce their impact on circuit performance and prevent system-level failures is crucial. The following sections will examine these mitigation strategies in detail.

\section{Aging Mitigation Techniques}
\label{mitigation}

\begin{figure*}
    \centering
    \includegraphics[width=\linewidth]{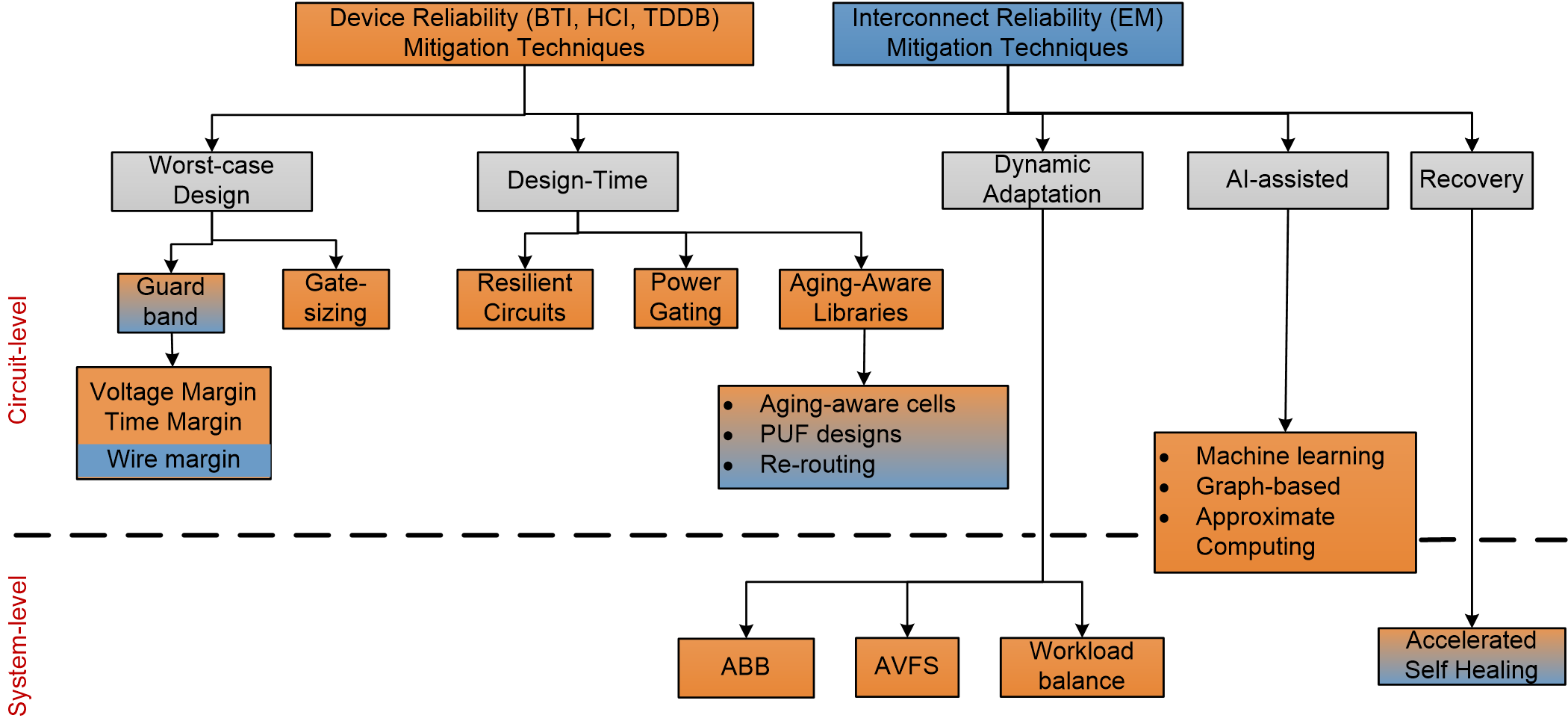}
    \caption{Taxonomy of device reliability (BTI, HCI, TDDB) and EM mitigation techniques for ICs at circuit-level and system-level.}
    \label{fig:taxonamy_ageMiti}
\end{figure*}

Aging mitigation techniques aim to reduce the impact of aging-related degradation in IC design, ensuring prolonged operational lifetimes, reliable performance, and lower failure rates. Over the last thirty years, various techniques have been developed to address reliability issues in devices and interconnects at both circuit and system levels. These techniques range from static, design-time solutions to dynamic, runtime approaches. As summarized in Figure ~\ref{fig:taxonamy_ageMiti}, the techniques are categorized into five primary groups, each comprising multiple techniques applicable at the circuit or system level ~\cite{guo2020_ICIntro_CircadianRhythmsFuture}. The subsequent sections will detail each mitigation technique from both circuit-level and system-level perspectives, supported by relevant literature demonstrating their effectiveness.
\subsection{Circuit-level Aging Mitigation Techniques}
Circuit-level aging mitigation techniques aim to directly counteract aging effects in circuit designs by improving the lifespan of critical components through strategic design adjustments and adaptive circuitry. This section will survey various mitigation approaches implemented in widely utilized circuits, including digital logic circuits, analog circuits, and SRAM memory blocks. These techniques range from worst-case design margins to innovative aging resilient digital circuits, adaptive bias adjustments. By integrating these strategies, designers can not only enhance the robustness of individual circuits but also improve the overall reliability of the system throughout its operational life.

\subsubsection{Mitigation Techniques in Digital Circuit:}
\begin{table}[t]
\centering
\caption{List of important references on aging mitigation techniques for digital circuits.}
\label{Table:ageMiti_digital_RelatedWork}
\resizebox{0.35\textwidth}{!}{
\begin{tabular}{c|cl}
\toprule
\textbf{Technique} & \textbf{References} \\
\midrule
\textbf{Resilient Circuits} &  ~\cite{raji2023_ageMiti_resiliCir, shah2018_ageMiti_resiliCir, Wu2019_ageMiti_resiCir_RFF, Ernst2003_ageMiti_resilCir_RFF, raji2021_ageMiti_digi_resiliCir,jafari2019_ageMiti_digi_resiliCir, qi2008_ageMiti_ABB_NBTIResilientCircuits}\\
\textbf{Guard Band} &  ~\cite{Kang2007_ageMiti_Guardband, Zhang2009_ageMiti_Guardband, henkel2016_ageMiti_Guardband, amrouch2017_ageMiti_Guardband, Amrouch2017a_ageMiti_Guardband, vanSanten2016_ageMiti_Guardband, amrouch2016_ageMiti_Guardband_approx}\\
\textbf{Gate Sizing} & ~\cite{wu2009_ageMiti_gateSizing, wu2011_ageMiti_gateSizing, chen2012_ageMiti_gateSizing, yang2007_ageMiti_gateSizing, xiong2024_ageMiti_gateSizing, xiong2025_ageMiti_gateSizing, aguiar2019_ageMiti_gatesizing, kiamehr2015a_ageMiti_gateSizing, khoshavi2014a_ageMiti_gateSizing,Abbas2017_ageMiti_software_DWL, barke2014_ageMiti_gateSizing_agingModel, wu2011_ageMiti_gateSizing_powergating}\\
\textbf{Power Gating} & ~\cite{wu2011_ageMiti_gateSizing_powergating, Bhattacharjee2021_ageMiti_digi, Chen2021_ageMiti_digi, liu2010_ageMiti_EvaluationPowerGating}\\
\bottomrule
\end{tabular}
}
\end{table}
Digital circuits that process binary signals are especially vulnerable to BTI and HCI. Both aging mechanisms cause an increase in the threshold voltage of MOSFETs, leading to reduced switching speeds and increased power consumption. Digital circuits, consisting of logic gates, combinational blocks, and flip-flops, are susceptible to timing errors resulting from aging-related delay degradation. Moreover, these timing errors caused by aging might lead to system-level malfunctions. Therefore, it is essential to design resilient digital logic circuits. Several mitigating strategies have been suggested in the literature to mitigate aging-induced degradation in digital circuits, and are summarized in Table ~\ref{Table:ageMiti_digital_RelatedWork}. 

\textbf{\textit{Resilient Circuit Design Techniques}}: 
Significant efforts have been made to redesign conventional basic digital circuit blocks to improve aging resilience and recovery. The most commonly used circuit elements include various logic gates, flip-flops, and the 6T SRAM structure. Numerous robust circuit design techniques for logic gates and flip-flops have been proposed to tackle aging-related issues. These novel designs aim to minimize aging effects in traditional logic gates, ensuring sustained reliability and performance over time. Although substituting elementary blocks with novel designs is challenging in contemporary design flows, it offers a radical solution for improving chip reliability.

The conventional inverter consists of an NMOS and a PMOS transistor, with the PMOS being particularly vulnerable to NBTI damage, which presents a greater threat than PBTI in MOSFETs. ~\cite{shah2018_ageMiti_resiliCir, Gupta2021_ageMiti_digi} proposed nMOS-only Schmitt trigger with voltage booster (NST-VB) circuits to improve NBTI tolerance and enhance soft error hardening. These designs replace the traditional p-MOS-based pull-up network with n-MOS transistors. In ~\cite{Morsali2022_ageMoni_RO}, a new inverter structure utilizing a switched pseudo current mirror to control the pull-up network and reduce short-circuit current during transitions demonstrated strong robustness against aging.

Beyond logic gates, aging mitigation in digital sequential blocks like flip-flops (FF) is also critical. Timing violations in flip-flops can compromise pipeline functionality. The Raji research group examined techniques to enhance the lifespan of various flip-flop designs ~\cite{jafari2019_ageMiti_digi_resiliCir, raji2021_ageMiti_digi_resiliCir, raji2023_ageMiti_resiliCir}. A reconfigured master-slave flip-flop was introduced to improve reliability by reducing stress on certain transistors within the feedback loop ~\cite{Jafari2020_ageMiti_digi_FFs}. This group also developed pulsed flip-flops with a modified pull-down network to reduce stress duration on pulsed clock transistors ~\cite{Jafari2021_ageMiti_digi}.

The RAZOR flip-flop (R-FF) was introduced for dynamic detection and correction of timing errors by adjusting the supply voltage ~\cite{Ernst2003_ageMiti_resilCir_RFF}. R-FF was later explored as a monitoring and reconfiguration technique to mitigate aging-related timing issues ~\cite{Wu2019_ageMiti_resiCir_RFF}. However, R-FFs require high-level system assistance, leading to performance penalties, commonly referred to as cycle loss ~\cite{Silva2017_ageMiti_digi}. To address this, Silva et al. proposed two alternatives: Time-Borrowing Flip-Flop (TBFF) and Alternative Path Activation Flip-Flop (APAFF) to mitigate aging-induced timing violations without cycle loss. TBFF allocates slack between critical and logical paths in the next sequential stage without requiring shadow cells, control cells, or additional circuitry, but it demands sufficient slack margin to accommodate delays. APAFF addresses critical paths with high fan-out delay constraints using a shadow flip-flop, thus eliminating the need for time borrowing ~\cite{Silva2017_ageMiti_digi}.

\textbf{\textit{Guard Band}}: 
The guard-band technique, or worst-case design, is a common approach in circuit design to safeguard against performance degradation due to aging mechanisms such as BTI, HCI, and TDDB. It involves adding a safety margin into the design to accommodate aging-induced fluctuations, ensuring circuits meet timing and performance requirements over their operating lifespan ~\cite{khoshavi2017_SurveyCMOSAging}. In digital circuits, two main guard bands are typically used to address aging concerns. The first is the \textit{voltage guard-band}, which sets the operating voltage marginally above the minimum required level. This allows the circuit to handle increased transistor threshold voltage caused by aging, thereby maintaining functionality and reliability. The second is the \textit{timing guard-band}, which introduces additional timing margins to accommodate delay shifts from aging. This is particularly critical for paths susceptible to timing violations that could cause functional failures. While guard-band techniques effectively mitigate aging effects, they introduce certain trade-offs. A higher voltage margin increases energy consumption throughout the circuit's lifetime ~\cite{Kang2007_ageMiti_Guardband, Zhang2009_ageMiti_Guardband}. Timing guard-bands often reduce circuit speed, as the added margins lower the maximum operating frequency and introduce performance overhead. Excessive use of timing guard-bands can also lead to increased power consumption, area, and speed penalties. The Amrouch research group analyzed guard-band techniques in various digital circuits and SRAM designs to mitigate aging and process-voltage-temperature (PVT) variations over the circuit's lifetime ~\cite{vanSanten2016_ageMiti_Guardband, henkel2016_ageMiti_Guardband, amrouch2016_ageMiti_Guardband_approx, amrouch2017_ageMiti_Guardband, Amrouch2017a_ageMiti_Guardband}. Designers must carefully balance efficiency and reliability when implementing guard-band techniques ~\cite{khoshavi2017_SurveyCMOSAging}.

\textbf{\textit{Gate Sizing}}:
Gate sizing is another worst-case design technique used to mitigate aging-related degradation in digital circuits, particularly in nanoscale CMOS technologies ~\cite{yang2007_ageMiti_gateSizing, chen2012_ageMiti_gateSizing, khoshavi2017_SurveyCMOSAging}. This technique targets critical gates that are highly susceptible to aging-induced delays, ensuring sustained performance over time. Adjusting transistor widths and gate dimensions reduces threshold voltage variations, delays, and thermal hotspots, thereby increasing circuit lifetime. Resizing gates along high-switching-activity paths mitigates NBTI/PBTI stress and enhances thermal resilience ~\cite{xiong2024_ageMiti_gateSizing, xiong2025_ageMiti_gateSizing}. Gate sizing is often combined with buffer insertion to reduce aging effects. Buffers in long interconnects mitigate signal latency and thermal stress, enhancing both durability and thermal resilience. Studies indicate that buffer insertion improves aging immunity ~\cite{xiong2024_ageMiti_gateSizing}. Additionally, combining gate sizing with power gating reduces leakage current and facilitates NBTI recovery during sleep modes ~\cite{khoshavi2014a_ageMiti_gateSizing, wu2011_ageMiti_gateSizing_powergating}.  

Although gate sizing improves reliability, excessively large transistors increase area and leakage power. Studies suggest that thermal immunity can be enhanced with minimal area overhead, but trade-offs in power and performance must be carefully considered ~\cite{aguiar2019_ageMiti_gatesizing}. Gate scaling remains a widely used technique for aging mitigation, but it faces challenges from process variation due to shrinking transistor dimensions. Researchers are exploring optimization strategies using predictive machine learning algorithms and integrating techniques such as buffer insertion and power gating to extend circuit lifespans under NBTI, PBTI, and HCI stress ~\cite{wu2011_ageMiti_gateSizing_powergating, khoshavi2014a_ageMiti_gateSizing, xiong2024_ageMiti_gateSizing, xiong2025_ageMiti_gateSizing}.

\begin{figure}[!t]
    \centering
    \includegraphics[width=0.95\linewidth]{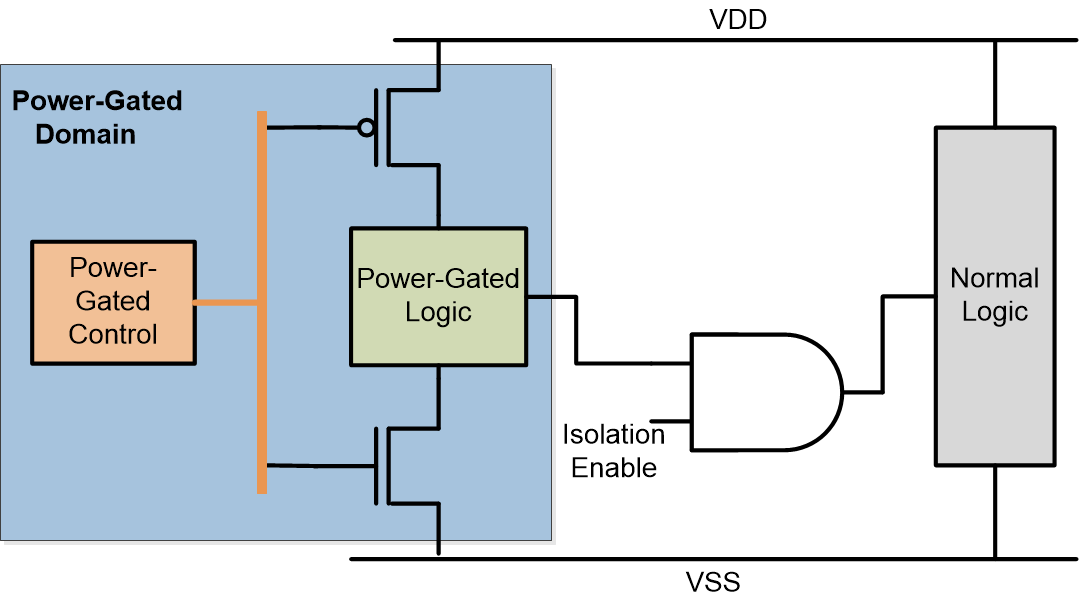}
    \caption{Illustration of power gating technique for aging mitigation.}
    \label{fig:powergating}
\end{figure}
\textbf{\textit{Power Gating}}:
Power gating is a widely used low-power design strategy that can also minimizes aging-related degradation in digital and SRAM circuits by disconnecting the power supply to inactive circuit sections. It reduces operating time and temperature, thereby mitigating aging effects and facilitating transistor recovery during sleep intervals ~\cite{khoshavi2014a_ageMiti_gateSizing}. This mitigation technique addresses three key aspects ~\cite{khoshavi2014a_ageMiti_gateSizing, liu2010_ageMiti_EvaluationPowerGating}: (i) minimizing active duration to limit stress exposure that accelerates BTI aging, (ii) reducing switching activity to lower power dissipation and thermally activated aging, and (iii) enabling MOSFET recovery from BTI-induced degradation during sleep mode. Figure ~\ref{fig:powergating} illustrates a standard power gating block diagram for integrated circuits. Power-gated logic connects to power rails via sleep transistors, which are controlled by a power-gated control logic unit. Increasing sleep transistor sizes minimizes voltage drop during activation, reducing stress on neighboring components ~\cite{khoshavi2014a_ageMiti_gateSizing}. An isolation enable signal separates the power-gated logic domain from the standard logic domain. 

Conventional power gate designs face two primary challenges. Activating sleep transistors simultaneously can induce a large surge current in the power supply network, causing IR drop and signal integrity issues. Alternatively, distributed sleep transistor networks (DSTNs) may experience BTI wearout, increasing wake-up delay. A reconfigurable circuit design with a novel aging-aware wake-up sequence was proposed to mitigate BTI effects ~\cite{Chen2021_ageMiti_digi}. An NBTI-aware power gating architecture that periodically activates and deactivates sleep transistors was introduced to enhance circuit durability, where additional sleep transistors engage when NBTI degradation reaches a threshold, allowing more recovery time and extending the lifespan of sleep transistors ~\cite{Bhattacharjee2021_ageMiti_digi}. In addition, power gating also presents limitations such as process variations, complex predictive models for optimizing sleep intervals, and challenges in sleep transistor sizing to minimize leakage current.

\renewcommand{\arraystretch}{2}
\begin{table}[t]
\centering
\caption{List of important references on aging Characterization and Mitigation techniques in SRAMs.}
\label{Table:ageMiti_SRAM_RelatedWork}
\resizebox{0.95\linewidth}{!}{
\begin{tabular}{c|c}
\toprule
\textbf{Classification} & \textbf{Important References}   \\
\midrule
\textbf{SRAM Cells} & \makecell{~\cite{Khan2014_BTI_SRAM, kumar2006a_ageMiti_ImpactNBTISRAM, Mostafa2011_ageMiti_SRAMs, Rauch07_BTIVari, Goel14_BTIMech_DCAC, Zuo2017_ageMiti_SRAM, Zuo2020_ageMiti_SRAM, Shah2018_ageMiti_SRAM, Listl2019_ageMiti_SRAM, listl2022_ageMiti_SRAM} \\
 ~\cite{ Kaczer10_BTIVari, Goel15_BTIVari_SRAM, JIN2018_BTI_SRAMvari, Naphade14_BTIVari_SRAM, Mishra2019_BTIVari_SRAM, Chenouf2020_ageMiti_SRAM, Shah2018a_ageMiti_SRAM, Sadeghi2018_ageMiti_SRAM, babu2018_SRAM_AnalyzingImpactNBTI, picardo2022_SRAM_IntegralImpactPVT, shaik2024_SRAMIMC_ImpactAgingProcess}}  \\

\textbf{\makecell{SRAM and its \\Peripheral Circuits}} & ~\cite{Menchaca2012_ageMiti_SRAM_Peri_SA, Agbo2013_ageMiti_SRAM_peri_SA, Agbo2015_ageMiti_SRAM_peri_WD, Agbo2019_ageMiti_SRAM_peri, Agbo2017_ageMiti_SRAM_peri_SA, Sadeghi2020_ageMiti_SRAM_peri, Kinseher2017_ageMiti_SRAM_peri, ceratti2014_ageMiti_SRAM_OnChipSensorMonitor, Kraak2020_age_Miti_SRAMperi, Agbo2018_ageMiti_SRAMPeri, Kraak2017_ageMiti_SRAMPeri, Kraak2019_ageMiti_SRAMPeri, Dounavi2019_ageMiti_SRAMPeri, Dounavi2020_ageMiti_SRAMPeri, Dounavi2021_ageMiti_SRAMPeri}   \\

\textbf{\makecell{SRAM (Cache) \\ Memory Systems}} &  ~\cite{Shin2008_ageMiti_SRAM_sys, Rohbani2019_ageMiti_SRAM_sys, Sharifi2020_ageMiti_Sys, Kang2007a_ageMiti_SRAM_IDDQ, alorda2016_ageMiti_SRAM_OnlineWriteMargin, ceratti2014_ageMiti_SRAM_OnChipSensorMonitor, haggag2007_ageMiti_SRAM_UnderstandingSRAM, wang2014_ageMiti_SRAM_SiliconOdometersCompact, abu_rahma2011_ageMiti_SRAM_CharacterizationSRAMSense, ahmed2016a_ageMiti_SRAM_OnlineMeasurementDegradation, saikia2021_SRAMIMC_ModelingDNN, shaik2024_SRAMIMC_ImpactAgingProcess, dilopoulou2023_SRAMIMC_BTIAging, tsai2012_ageMiti_EmbeddedSRAMRing, Halak2020_ageICbook}  \\   
\bottomrule                   
\end{tabular}
}
\end{table}

\subsubsection{Aging Mitigation Techniques in SRAM and its Peripherals}
SRAM serves as a significant component in modern processors and is critical to their performance characteristics ~\cite{Molina2003_SRAM_cacheSpace}. Consequently, SRAM reliability is essential for the reliable operation of modern systems. In sub-nanometer technologies, SRAM reliability has been severely impacted by increased process variability and aging mechanisms, including BTI and HCI. BTI degradation accelerates under extreme stress and elevated temperatures ~\cite{Khan2014_BTI_SRAM, Mishra2019_BTIVari_SRAM}. The data stored in SRAM cells influences the transistor aging rate ~\cite{shaik2020_SRAM_AnalysisSRAMMetrics, Goel15_BTIVari_SRAM}. In back-to-back connected inverters within an SRAM cell, MOSFETs experience BTI stress even during hold operations. Consequently, SRAM cells remain susceptible to BTI degradation regardless of the stored data. The degradation in $V_{TH}$ affects the stability of SRAM (read, write, and hold), which is typically assessed using static noise margin (SNM). Therefore, developing effective aging monitoring and mitigation techniques for SRAM is essential to predict potential failures during the memory lifetime and enable timely circuit repair to ensure reliable long-term operation. Table ~\ref{Table:ageMiti_SRAM_RelatedWork} summarizes the key references that address the aging issues in SRAMs, peripheral circuits and the SRAM-based systems. Details of each category is discussed below.


\textbf{\textit{SRAM Characterization and Mitigation: }} 
Numerous works have characterized the read stability and write-ability of SRAM using static and dynamic metrics ~\cite{Khan2014_BTI_SRAM, picardo2022_SRAM_IntegralImpactPVT, babu2018_SRAM_AnalyzingImpactNBTI, chaudhari2018_SRAM_CorrelationDynamicStatic}. With advancements in technology, process variations and aging-induced variations have received increased attention. Recent studies have examined the combined impact of process variations and BTI on SRAM stability ~\cite{Goel15_BTIVari_SRAM, Naphade14_BTIVari_SRAM, Mishra2019_BTIVari_SRAM, MUKHOPADHYAY2018_BTIVari_DS}. ~\cite{JIN2018_BTI_SRAMvari} analyzes the impact of transistor-level BTI degradation on circuits, focusing on ROs and SRAM. It demonstrates robust 10 nm SRAM and product-level HTOL reliability up to 500 hours, highlighting the importance of accurately characterizing BTI-induced $V_{TH}$ variability. Ref. ~\cite{shaik2022_SRAM_Vari_FinFET} investigates the impact of NBTI variations on SRAM stability in FinFET technology.


\textbf{\textit{SRAM and its Peripheral Circuits Characterization and Mitigation: }}
Current research indicates that MOSFET degradation in SRAM cells and peripheral circuits (such as sense amplifiers, write drivers, and decoders) leads to a gradual decrease in speed performance. The input offset voltage of sense amplifiers (SA) rises with MOSFET degradation, increasing the likelihood of failures during SRAM operation. To the best of our knowledge, limited studies have focused on the effect of BTI on sense amplifiers and write drivers ~\cite{Agbo2019_ageMiti_SRAM_peri, Agbo2013_ageMiti_SRAM_peri_SA, Agbo2017_ageMiti_SRAM_peri_SA, Agbo2015_ageMiti_SRAM_peri_WD, Kraak2017_ageMiti_SRAMPeri, shaik2020_SRAM_AnalysisSRAMMetrics}. Several works have characterized aging effects and proposed mitigation techniques for peripheral circuits in SRAM, particularly evaluating their impact on read and write path delays. The degradation of the read path is more severe than the write path, with sense amplifiers being the most susceptible circuit component ~\cite{Agbo2018_ageMiti_SRAMPeri}. Research from TUDelft has investigated BTI effects on key peripheral circuits and proposed mitigation strategies for improving read and write paths in embedded memories ~\cite{Kraak2017_ageMiti_SRAMPeri, Kraak2019_ageMiti_SRAMPeri, Kraak2020_age_Miti_SRAMperi, Agbo2018_ageMiti_SRAMPeri, Agbo2019_ageMiti_SRAM_peri}. 

A technique for characterizing SRAM sense amplifier input offset voltage for yield prediction is presented in ~\cite{abu_rahma2011_ageMiti_SRAM_CharacterizationSRAMSense}. It employs two resistor-string 6-bit digital-to-analog converters (R-DACs) to generate various bit-line voltage differences required for offset voltage estimation. ~\cite{Kinseher2017_ageMiti_SRAM_peri} examines transistor aging in SRAM peripheral logic in modern System-on-Chips (SoCs), revealing that peripheral degradation reduces access speed and read margin while improving write margin. The study underscores the importance of analyzing aging mechanisms at the system level rather than in individual sub-circuits.

\textbf{\textit{Aging Monitoring and Mitigation in SRAM-based Systems: }}
Various techniques have been proposed to monitor aging-related degradation in SRAM systems, but challenges remain, such as difficulty in locating defective cells, high silicon area costs, and design complexity. In ~\cite{Kang2007a_ageMiti_SRAM_IDDQ}, IDDQ current is used to characterize NBTI-induced degradation, but it measures overall leakage without identifying specific defective cells. An on-chip NBTI sensor for offline monitoring of SRAM writes was proposed in ~\cite{wang2014_ageMiti_SRAM_SiliconOdometersCompact}, assuming power gating and a non-aged reference column. Monitoring solutions based on embedded ring oscillators and voltage-controlled oscillators have also been explored ~\cite{tsai2012_ageMiti_EmbeddedSRAMRing}. A write margin degradation technique using wordline write trip voltage (WWTV) was introduced in ~\cite{alorda2016_ageMiti_SRAM_OnlineWriteMargin} but requires multiple voltage levels. Error-correcting codes (ECCs) ~\cite{haggag2007_ageMiti_SRAM_UnderstandingSRAM} can prevent aging-related failures but require redundancy as aged cells increase. A BTI degradation estimation method using pseudo write and read operations was proposed in ~\cite{ahmed2016a_ageMiti_SRAM_OnlineMeasurementDegradation}. ~\cite{Shah2018_ageMiti_SRAM} proposed an on-chip adaptive body bias (O-ABB) circuit to compensate for NBTI aging and improve yield, showing reduced NBTI impact on SRAM stability. ~\cite{masoumian2022_ageMiti_SRAMs} demonstrated the reliability of SRAM Physical Unclonable Functions (PUFs) in FinFET nodes, with minimal noise and temperature sensitivity across 16nm, 14nm, and 7nm nodes.

Periodic refresh strategies have also been explored. ~\cite{kumar2006a_ageMiti_ImpactNBTISRAM} suggested periodic data flipping and an input switching sense amplifier to reduce degradation. ~\cite{Dounavi2020_ageMiti_SRAMPeri,Dounavi2019_ageMiti_SRAMPeri,Dounavi2021_ageMiti_SRAMPeri} proposed an aging sense and repair framework for early detection and recovery. ~\cite{Listl2019_ageMiti_SRAM,listl2022_ageMiti_SRAM} developed a workload-aware analysis framework for SRAMs, predicting sense amplifier degradation and proposing Mitigation of AGIng Circuitry (MAGIC) to reduce degradation by 26\% and extend lifespan by 3$\times$. While ~\cite{Zuo2017_ageMiti_SRAM} introduced asymmetric scaling for SRAM cells to improve aging quality retention and optimize lifetime yield, integrating it into the SRAM compiler CACTI ~\cite{murali2009_cacti_SRAMcompiler}.

~\cite{Sadeghi2018_ageMiti_SRAM} proposed swapping stored values in caches to reduce stress and extend memory lifespan by 1.91$\times$. ~\cite{Yang2018_ageMoni_review} proposed monitoring and mitigating transistor degradation using in situ and in-field methods with recovery vectors. ~\cite{Ghaderi2018_ageMiti_FPGA} developed a stress-aware approach (named STABLE) to mitigate Static Noise Margin (SNM) degradation in SRAM cells, increasing reliability in FPGAs. ~\cite{Gabbay2021_ageMiti_sys_asymAge} introduced an asymmetric aging-aware microarchitecture for execution units, register files, and memory hierarchy, significantly reducing asymmetric aging stress. Accelerated aging mechanisms like BTI degrade memory cell noise margins, increasing failure rates. Error-correction codes, such as Hamming codes, help mitigate these issues. A novel coding approach reduced SNM degradation by 45.2\% by balancing stored signal probabilities ~\cite{Golanbari2017_ageMiti_SRAM}.

\textbf{\textit{Aging Characterization on SRAM-based In-Memory Computing (IMC) Architectures:}} SRAM-based in-memory computing is an emerging domain concentrating on multiple applications, including edge computing, machine learning, and artificial intelligence ~\cite{MITTAL2021_SRAMIMC_survey, saikia2021_SRAMIMC_ModelingDNN, xu2022_SRAMIMC_AI, Runxi2023_SRAMIMC_edgecomputing, Runxi2024_SRAMIMC_accelaration}.  Very limited studies have been carried out on the impact of aging on IMC architecture. 
~\cite{dilopoulou2023_SRAMIMC_BTIAging} examines the effects of BTI aging on SRAM-based IMC architectures and proposes a mitigation approach that involves frequently switching between active and inactive transistors in these cells to enhance their reliability and longevity. ~\cite{shaik2024_SRAMIMC_ImpactAgingProcess} presents a comprehensive examination of the effects of aging and process variability on commonly used SRAM-based 6T and 8T IMC architectures. It analyzes the stochastic behavior of degradation in the two IMC architectures and identifies the worst-case failures. This research is among the initial studies to present an examination of BTI-induced aging across various IMC logic operations, offering insights for the design of resilient SRAM-based IMC architectures. Given that SRAM and its peripheral circuits are integral to IMC design, the aforementioned aging mitigation techniques have to be employed in order to develop aging-aware IMC architectures that are suitable for various applications. 

\subsubsection{Aging Mitigation Techniques in Analog Circuits}
Similar to digital circuits, analog circuits, which handle continuous signals, are also susceptible to aging phenomena such as BTI, HCI, and TDDB, leading to performance degradation in metrics like offset, duty cycle, and gain. Assessing post-degradation performance is complex, posing significant challenges for analog designers. This necessitates comprehensive simulation methodologies to predict functional and performance degradation during the pre-silicon phase. Despite limited progress, some reliability simulation approaches have been developed to predict aging degradation in modern analog circuits ~\cite{Maricau13_analog_rel,art_AnalogRelSim_ageMOS}. As summarized in Table ~\ref{tab:analogagingmitigation}, this section surveys design strategies to mitigate aging-related degradation in analog circuits.

Since the 1970s, researchers have studied device degradation models, but the impact of aging on analog circuits became apparent in the early 2000s due to transistor scaling, which increased susceptibility to degradation ~\cite{jha2005_age_analog_NBTIDegradationIts}. This underscores the importance of considering aging effects during early design stages. Early studies focused on standard analog circuits such as comparators, ADCs, DACs, and voltage regulators ~\cite{gielen2011_age_Analog_AnalogCircuitReliability, mahato2013_age_analog_ImpactTransistorAging}. Growing interest in non-Von Neumann architectures, including in-memory and neuromorphic computing, has also led to investigations into BTI and HCI effects on neuromorphic and SRAM-based IMC circuits. The electronics industry faces challenges in assessing reliability impacts on ICs. While existing reliability EDA tools address process-related issues, they are less effective for aging-related degradation in analog circuits. Several simulation methodologies have been developed to evaluate aging effects. ~\cite{Yan2009_age_Analog} presents a reliable simulation methodology for high-speed Flash ADCs, identifying NBTI as the primary failure mechanism under elevated temperatures. A reliability simulation framework from device to circuit level has been used to study BTI and HCI effects on analog circuits such as comparators, VCOs, and DACs. Another reliability tool, `CASE,' developed using MATLAB, has been used to analyze aging in operational amplifiers and ring oscillators. ~\cite{afacan2019_age_Analog_miti} proposes an aging-aware analog circuit sizing tool to improve design automation, while ~\cite{kuntman2019_age_analog_ReliabilityEstimation} presents a statistical modeling approach for transistor degradation, showing high accuracy and short simulation times.

Aging mitigation strategies have also been explored in analog circuits. Guardbanding techniques ensure circuit performance under worst-case aging. An extension of the gm/ID sizing method optimizes fresh circuit designs for reliability by considering aged transistor parameters ~\cite{hellwege2013_age_analog_UsingOperatingPointdependent}. Efficient sizing methods have been proposed to balance lifetime and layout area, revealing that extending lifetime requires additional layout area ~\cite{pan2010_age_analog_ReliabilityAnalysisAnalog, pan2012_age_analog_ReliabilityOptimizationAnalog}. Reliability-aware analog circuit design is typically divided into two approaches: (i) precautionary — designing with aging effects in mind; and (ii) sense-and-react — monitoring performance degradation and initiating corrective mechanisms ~\cite{afacan2017_age_Analog_review}. Aging-aware design strategies have been proposed for common analog circuits such as operational amplifiers, ADCs, DACs, comparators, and VCOs, emphasizing BTI, HCI, and OPDD mitigation ~\cite{melikyan2024_age_analog_AgingAwareDesignMethod}. Comprehensive design strategies for aging mitigation have been suggested, including low-voltage design, power gating, regulating device voltages, current trimming, duty-cycle correction, and offset cancellation ~\cite{yellepeddi2020_age_analog_AnalogCircuitDesign}. Techniques such as BTI-clocking and slowing down slew rates have also been proposed. Additionally, diode-connected transistors have been used to mitigate BTI and HCI effects in silicon neurons and neuromorphic systems ~\cite{shaik2024_age_analog_ReliabilityawareDesignSiN, shaik2022_age_analog_ReliabilityawareDesignTemporal, picardo2022_age_analog_EnablingEfficientRate}. These strategies form the foundation for aging-aware analog circuit design.

\begin{table*}[!t]
\centering
\caption{Literature on aging characterization and mitigation techniques for various analog CMOS circuits.}
\label{tab:analogagingmitigation}
\resizebox{0.75\textwidth}{!}{
\begin{tabular}{c|c|c|c|c}
\toprule
\textbf{References} & \textbf{Aging Mechanisms}  & \textbf{Analog Circuits} & \textbf{\makecell{Aging \\ Characterization}} &\textbf{\makecell{Aging \\Mitigation}} \\

\midrule
~\cite{Yan2009_age_Analog, yan2008_age_analog_ReliabilitySimulationDesign} & NBTI, HCI, TDDB & Comparator, flash ADC  & \checkmark & \- \\

~\cite{Afacan2015_age_analog_relTool, afacan2019_age_Analog_miti, afacan2017_age_Analog_review} & NBTI, HCI  & \makecell{Operational \\trans-conductance (OTA)}  & \checkmark & \checkmark \\

~\cite{gielen2010_age_analog_DesignAutomationReliable, gielen2011_age_Analog_AnalogCircuitReliability, gielen2012_age_analog_DesigningReliableAnalog} & BTI, HCI, TDDB  & LC-VCO & \checkmark & \checkmark \\

~\cite{gielen2008_age_analog_EmergingYieldReliability} & TDDB, HCI, NBTI  & DAC &   & \checkmark  \\

~\cite{habal2013_age_analog_EvaluatingAnalogCircuit} & BTI  & Differential amplifier & \checkmark & \\

~\cite{hellwege2013_age_analog_UsingOperatingPointdependent, Hellwege2015_ageMiti_analog_gmID, hillebrand2016_age_analog_StochasticLUTbasedReliabilityaware} & NBTI, HCI & \makecell{Common source amplifier\\ two-stage Op-amp} & \checkmark & \checkmark\\

~\cite{jha2005_age_analog_NBTIDegradationIts} &  NBTI & CMOS Op-amp & \checkmark & \\

~\cite{ho2014_age_analog_ImpactHotCarrier} & HCI  & LC-VCO & \checkmark & \\

~\cite{mahato2013_age_analog_ImpactTransistorAging} &  HCI, TDDB, BTI & Cascode \& Folded LNA & \checkmark & \\

~\cite{mahat2012_age_analog_oOffsetMeasurementMethod} &  BTI & OTA & \makecell{\checkmark} & \\

~\cite{maricau2011_age_analog_TransistorAginginduced, maricau2011_age_analog_ComputerAidedAnalogCircuit} & HCI, BTI, TDDB  & \makecell{LC-VCO, OTA, \\ comparator, current mirror} &  \checkmark & \checkmark \\

~\cite{pan2010_age_analog_ReliabilityAnalysisAnalog, pan2012_age_analog_ReliabilityOptimizationAnalog} &  NBTI, HCI, & miller OTA & \checkmark & \\

~\cite{more2011_age_analog_ReliabilityAnalysisBuffer} & BTI, HCI  & buffer, Successive ADC & \checkmark & \\

~\cite{Prabhu2019_age_Analog} & NBTI, PBTI  & Comparator & \checkmark & \\

~\cite{sanic2018_age_analog_TimedependentDielectricBreakdown} & TDDB  & op-amp, RF-mixer & \checkmark & \\

~\cite{simicic2019_age_analog_UnderstandingImpactTimedependent} &  BTI-induced variability & comparator & \checkmark & \\

~\cite{dhar2023_age_analog_AgingModelCurrent} & HCI & current DAC & \checkmark & \checkmark \\

~\cite{park2023_age_analog_TemperatureAgingCompensatedRC} &  EM, HCI & RC-Oscillator & \checkmark & \checkmark\\

~\cite{shaik2022_age_analog_ReliabilityawareDesignTemporal, shaik2024_age_analog_ReliabilityawareDesignSiN, picardo2022_age_analog_EnablingEfficientRate, Krishna2024_age_analog_synapse} &  BTI, HCI & \makecell{Silicon Neuron (SiN),\\ Silicon Synapse (SiS)}  & \checkmark & \checkmark \\

\bottomrule
\end{tabular}
}
\end{table*}

\subsection{System-level Aging Mitigation Techniques}

Thus far, we have discussed aging mitigation approaches at the circuit level. However, circuit-level techniques typically incur additional area or power overhead. To address this, many system-level approaches have been proposed, which build upon existing infrastructure and circuit components. These techniques leverage trade-offs introduced by single-layer optimization, aiming to achieve more efficient and scalable aging mitigation without significantly compromising overall system performance. Table ~\ref{Table:ageMiti_system_RelatedWork} summarizes the important references demonstrated the system-level mitigation techniques. It is important to understand the emergent of mitigation techniques couple with existing techniques accelerate the mitigation of aging effects in ICs. 

\begin{figure}[!t]
    \centering
    \includegraphics[width=0.95\linewidth]{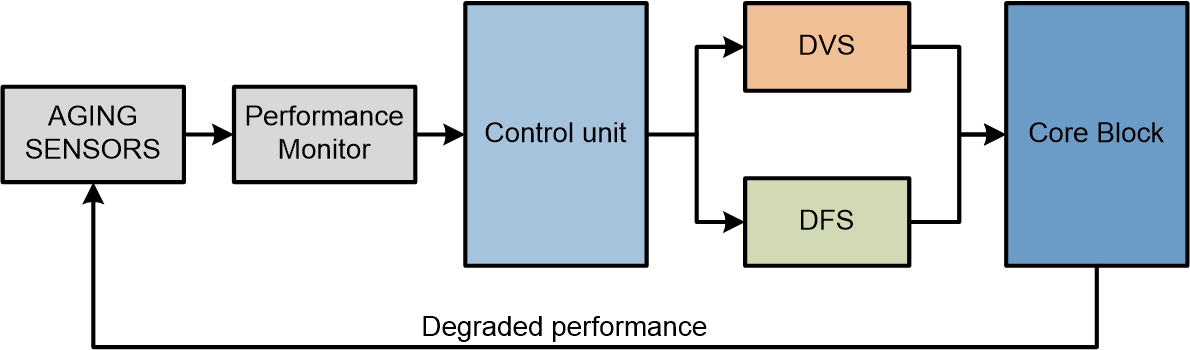}
    \caption{Conceptual block‐diagram of adaptive voltage and frequency scaling (AVFS) to mitigate PVT and aging in integrated circuits.}
    \label{fig:AVFS_diag}
\end{figure}

\textbf{\textit{Adaptive Voltage and Frequency Scaling (AVFS):}} 
AVFS dynamically adjusts the supply voltage and operating frequency based on real-time assessments of circuit conditions to ensure optimal functionality and minimize aging-induced degradation. It integrates dynamic voltage scaling (DVS) and dynamic frequency scaling (DFS), both initially developed for energy-efficient circuit design. Figure ~\ref{fig:AVFS_diag} illustrates the conceptual flow of AVFS, where real-time data, including the propagation delay of critical paths, is fed to control logic to modulate the system’s voltage or operating frequency. The cycle repeats, incorporating feedback on performance degradation from embedded aging sensors to refine adjustments. This strategy was widely adopted in the late 2000s to reduce the existing system guard band ~\cite{tschanz2007_ageMiti_AVFS, kumar2009b_ageMiti_AVFS}. However, AVFS has several limitations: (i) real-time AVFS control logic requires complex algorithms and robust aging models (including device scaling) to predict long-term performance degradation, increasing design complexity and area overhead; (ii) temperature fluctuations can intensify aging effects, complicating AVFS adjustments and potentially destabilizing system performance.

Recent research has focused on AVFS to mitigate process variations and aging-related degradation ~\cite{vincent2012_ageMiti_AVFS}. Adaptive methods in 90nm CMOS technology have been introduced to address performance degradation caused by aging, thereby reducing the aging guardband requirement ~\cite{tschanz2007_ageMiti_AVFS}. Robust circuits using in-situ timing monitoring can eliminate excessive time margins induced by process, voltage, and temperature (PVT) variations and aging effects ~\cite{shan2017_ageMiti_AVFS, shan2020a_ageMiti_AVFS}. The need for additional timing margins during signoff in aging-aware standard-cell timing libraries has been analyzed, and two algorithms have been proposed to minimize power and area penalties caused by over- or under-estimation of aging ~\cite{chan2014_ageMiti_AVFS}. Frequency reduction can prevent timing violations when aging slows down the critical path, and supply voltage adjustments can counteract performance deterioration ~\cite{scarpato2017_ageMiti_AVFS_thesis}. A guard-banding technique that employs adaptive supply voltage and body biasing to restore performance has been proposed, eliminating the need for large static guard bands ~\cite{kumar2009b_ageMiti_AVFS}. A comprehensive framework and control strategy have been introduced to enhance dynamic control of self-tuning parameters in digital systems under aging, improving computational efficiency and performance over the system’s lifespan, as validated through simulation results ~\cite{Mintarno2011ageMiti_digi}.

\textbf{\textit{Adaptive Body Bias Voltage (ABB): }} ABB is another dynamic technique for mitigating aging in integrated circuits that dynamically modifies the substrate (body) bias of MOS transistors to prevent degradation caused by BTI and HCI ~\cite{kumar2009b_ageMiti_AVFS, sivadasan2018_ageMiti_ABB_NBTIAgedCell, qi2008_ageMiti_ABB_NBTIResilientCircuits}. The ABB mitigation technique operates based on two primary principles: (i) Forward Body Bias (FBB), which decreases the $V_{TH}$ value, and (ii) Reverse Body Bias (RBB), which increases the $V_{TH}$ value. Initial studies on body biasing concentrated on power management; however, subsequent researchers explored the prospects of adaptive body biasing for the purpose of mitigating aging ~\cite{kumar2009b_ageMiti_AVFS}. Numerous studies have shown that ABB can substantially mitigate the degradation rate of transistor performance by offsetting the threshold voltage shifts linked to BTI ~\cite{sivadasan2018_ageMiti_ABB_NBTIAgedCell, qi2008_ageMiti_ABB_NBTIResilientCircuits, mostafa2012a_ageMiti_ABB_NBTIProcessVariations, blagojevic2016_ageMiti_ABB_FastFlexiblePositive, mhira2018_ageMiti_ABB_ResilientAutomotiveProducts}. 

Recent improvements have also incorporated adaptive body biasing (ABB) with on-chip feedback systems, facilitating real-time adjustments ~\cite{olivieri2005_ABB_NovelYieldOptimization}. A novel approach has been introduced for enhancing the yield of digital integrated circuits, addressing limitations in speed, dynamic power, and leakage power by employing process parameter estimation circuits and an integrated digital controller to regulate body bias. A minimal area overhead ABB circuit has been proposed to enhance system reliability and yield in submicrometer high-speed applications and microprocessors. The ABB circuit includes a threshold voltage-sensing circuit and an integrated analog controller, efficiently mitigating negative-bias temperature instability (NBTI) and process fluctuations to improve timing yield and overall yield ~\cite{mostafa2012a_ageMiti_ABB_NBTIProcessVariations}. Back biasing has also been identified as a critical factor in path selection for FDSOI circuits, which are increasingly employed for aging mitigation ~\cite{sivadasan2018_ageMiti_ABB_NBTIAgedCell}.

Moreover, mixed approaches that combine ABB with other aging mitigation techniques, such as dynamic voltage scaling (DVS) and optimized gate sizing, have been proposed to further enhance circuit robustness ~\cite{yan2003_ageMiti_ABB_DVS_CombinedDynamicVoltage, kumar2009b_ageMiti_AVFS}. Future strategies that integrate optimal frameworks and machine learning models for predictive control of body biasing are expected to enhance the precision and efficacy of aging compensation ~\cite{mintarno2010_ageMiti_ABB_OptimizedSelftuningCircuit}. ABB has been widely adopted in both digital circuits and SRAM (cache) memory systems. However, ABB has certain limitations, including (i) design complexities due to additional circuitry for bias generation and control logic and (ii) process variations that may impact ABB efficiency, necessitating resilient design methodologies to ensure reliability ~\cite{mostafa2012a_ageMiti_ABB_NBTIProcessVariations, narendra1999_ABB_PVT, chen2003_ABB_PVT_ComparisonAdaptiveBody}.

\textbf{\textit{Dynamic Workload Balancing: }} Dynamic Workload balance mitigates aging by redistributing computational stress to prevent overuse of degraded components. This technique involves in adaptive task scheduling that shifts tasks from degraded cores to healthier ones, Thermal-aware scheduling which prevents localized overheating and exacerbates aging in embedded systems. 

Several studies have explored aging mitigation at the system level. Workload classification has been introduced to mitigate aging. One study classified workloads into integer-majority and floating-point-majority categories based on instruction rates. By modifying the operating system scheduler to initiate recovery phases at appropriate times, the aging rate of a multicore processor was reduced by about 35\% over ten years of operation \cite{Sharifi2020_ageMiti_Sys}. Another approach proposed a static job scheduling technique for real-time MPSoC systems to concurrently optimize soft-error reliability (SER) and lifetime reliability (LTR) using evolutionary operators integrated with NSGAII and SPEA2. Experimental results demonstrated increased hypervolume and up to hundreds of times greater efficiency than leading methods \cite{Zhou2019_ageMiti_sys}. Task allocation strategies have also been explored. An adaptive task allocation method for multi-core embedded real-time systems (TAMER) was introduced, which optimizes core utilization and internal unit activity to reduce temperature variations and aging impact \cite{Saadatmand2021_ageMiti_embSys}. Another study proposed a low-overhead workload management technique for embedded GPUs that accounts for process variation and aging conditions, allocating distinct instructions to clusters to mitigate the aging impact. This approach extended GPU lifetime in more than 95\% of cases and reduced performance overhead by 40\% compared to standard compiler-based methods \cite{Lee2018_ageMiti_sys}. A task parallelism framework for multi-core systems has been introduced to address NBTI-related aging effects. This framework prioritizes activities based on their criticality, determines the optimal degree of parallelism (DOP) without time violations, and employs a task-to-core mapping technique. Results show that this approach can extend system longevity by a factor of 3.7 compared to the MAX DOP technique \cite{Chen2020_ageMiti_sys}. Stress-aware mapping strategies have also been proposed. A stress-aware loop mapping approach integrates intra-kernel and inter-kernel stress optimization techniques in the early stages of Coarse-Grained Reconfigurable Architectures (CGRAs). This method increases the mean time to failure (MTTF) by an average of 340.3\% and enhances stress reduction by up to 78.9\% \cite{Gu2017_ageMiti_digi}. Along this direction, \cite{Vijayan18_ageMiti_digiCir} presents a runtime monitoring and actuation method to alert timing-critical flip-flops of severe S-BTI stress, hence reducing monitoring costs by selecting representative offline flip-flops and performing online monitoring during static aging phases.  Experiments conducted on two processors established that 0.5\% of the total flip-flops are sufficient to be designated as representative flip-flops for S-BTI stress monitoring.  A low-overhead mitigation strategy is also suggested to reduce considerable flip-flops.

Besides, redundancy techniques have been explored. A time-redundant technique was introduced to counteract NBTI/PBTI aging effects on a processor's functional units. It identifies that half of the circuit can be NBTI-stressed during application execution, and the proposed solution doubles the system's lifetime by utilizing idle time effectively \cite{Abbas2017_ageMiti_software_DWL}. Finally, architectural improvements have been introduced to mitigate aging. A uni-directional shift register for on-chip digital low-dropout voltage regulators (LDOs) was developed to reduce NBTI effects and distribute electrical stress without adding power or area overhead. Simulations on an IBM POWER8 CPU showed a 43.2\% performance improvement over conventional designs, highlighting the reliability benefits of this architecture \cite{Wang2018_ageMiti_digi}.

\textbf{\textit{Accelerated Self-Healing:}} Accelerated self-healing (ASH) techniques have emerged as a proactive approach to mitigate aging-induced degradation and restore system functionality. This technique induces recovery phases, often during idle periods or through controlled circuit interventions, to reverse or slow down degradation. Several aging mechanisms (such as BTI and EM) consist of both reversible and irreversible components, where the irreversible component may partially revert under specific conditions. A biologically inspired approach has been proposed to mitigate irreversible degradation, reducing design margins and enhancing performance within a decade-long lifespan. This methodology has been demonstrated on commercial FPGAs \cite{Guo16_ASH_workHard}. A follow-up study introduced a method to activate and expedite recovery for both BTI and EM mechanisms through in-time scheduled recovery, effectively eradicating permanent wearout and introducing an additional design dimension \cite{Guo2017a_ASH_DeepHealing}. Research also demonstrated that adjusting the ratio of sleep to active modes can periodically rejuvenate systems, enhancing performance \cite{guo2014_ASH_modelDemo}. Similarly, an aging-aware placement technique for FPGA-based runtime reconfigurable architectures was introduced to optimize stress distribution and logical location during operation, reducing maximum stress and improving Mean Time To Failure (MTTF) \cite{zhang2015_ASH_STRAPStressAware}. 

Proactive scheduling for self-healing has also been explored \cite{morgul2022_ASH_SchedulingActiveAccelerated}. A global scheduler was proposed to perform scheduled recovery comprehensively, targeting the underlying causes of aging rather than just treating the symptoms. Experimental results demonstrated improved system longevity and stability. Recent work focuses on developing scalable ASH techniques to mitigate wearout effects in advanced technology nodes \cite{guo2017_ASH_DeepHealingEase, guo2020_ICIntro_CircadianRhythmsFuture, guo2022_ASH_designForRecovery}. Virtual Reconfigurable Circuits (VRC) have also been applied to fault mitigation by reconfiguring circuits to recover from faults such as Single Event Upsets (SEUs) \cite{Depanjali_2024_ASH_SEU_MissionCritical}. This method has demonstrated a substantial decrease in hardware usage and an 87\% enhancement in convergence performance for fault recovery in mission-critical applications, highlighting the potential for real-time fault recovery and system resilience.

ASH has recently been extended to flash memories, where proactive recovery techniques—known as the flash's Circadian Rhythm—have shown significant improvements. Studies demonstrated that early recovery can increase flash memory endurance by up to nine times and double its sustainability \cite{morgul2024unveiling, morgul2022towards}. These findings suggest that accelerated self-healing can be adapted to various memory and logic systems, enhancing long-term reliability and performance.

\renewcommand{\arraystretch}{1.5}
\begin{table}[t]
\centering
\caption{List of important references on aging Mitigation techniques at system level}
\label{Table:ageMiti_system_RelatedWork}
\resizebox{0.95\columnwidth}{!}{
\begin{tabular}{c|cl}
\toprule
\textbf{Mitigation Technique} & \textbf{References} \\
\midrule
\textbf{AVFS} & ~\cite{chan2014_ageMiti_AVFS,shan2017_ageMiti_AVFS,shan2020a_ageMiti_AVFS, kumar2009b_ageMiti_AVFS, tschanz2007_ageMiti_AVFS, vincent2012_ageMiti_AVFS,scarpato2017_ageMiti_AVFS_thesis}\\
\textbf{ABB} & ~\cite{narendra1999_ABB_PVT, sivadasan2018_ageMiti_ABB_NBTIAgedCell, chen2003_ABB_PVT_ComparisonAdaptiveBody, olivieri2005_ABB_NovelYieldOptimization, qi2008_ageMiti_ABB_NBTIResilientCircuits, mostafa2012a_ageMiti_ABB_NBTIProcessVariations, yan2003_ageMiti_ABB_DVS_CombinedDynamicVoltage, blagojevic2016_ageMiti_ABB_FastFlexiblePositive, mhira2018_ageMiti_ABB_ResilientAutomotiveProducts, mintarno2010_ageMiti_ABB_OptimizedSelftuningCircuit, milutinovic2009_ageMiti_ABB_DynamicVoltageFrequencyABB}\\
\textbf{Workload Balance} & ~\cite{Lee2018_ageMiti_sys, Chen2020_ageMiti_sys, Zhou2019_ageMiti_sys, Zhang2020_ageMiti_sys, Moghaddam2018_ageMiti_sys, Sangodoyin2021_ageMiti_sys, Gabbay2021_ageMiti_sys_asymAge, Sharifi2020_ageMiti_Sys, gabbay2023_ageMiti_Sys_EM, Moghaddasi2020_ageMiti_Sys, Shin2008_ageMiti_SRAM_sys, Rohbani2019_ageMiti_SRAM_sys, Saadatmand2021_ageMiti_embSys}\\
\textbf{Accelerated Self-Healing} & ~\cite{guo2014_ASH_modelDemo, Guo16_ASH_workHard, guo2017_ASH_DeepHealingEase, guo2022_ASH_designForRecovery, Guo2017b_ASH_ImplicationsCLASH, zhang2015_ASH_STRAPStressAware, Depanjali_2024_ASH_SEU_MissionCritical, morgul2022_ASH_SchedulingActiveAccelerated}\\
\bottomrule
\end{tabular}
}
\end{table}

\section{Software-based Aging Characterization and Mitigation Techniques}
\label{software}
This section discusses software-level strategies involving hardware-software co-design, leveraging machine learning and data-driven approaches to counteract aging effects.

\subsection{AI-Driven Approaches for Circuit Reliability}
Over the past two decades, artificial intelligence techniques, particularly machine learning and graph-based learning algorithms, have gained significant attention across various domains. Recent research has increasingly focused on applying these approaches to estimate and predict aging-related circuit performance degradation. The following subsections explore recent advancements in these domains.

\subsubsection{Machine Learning-based Aging Prediction}
Machine learning-based approaches have gained traction for aging-aware design and mitigation in integrated circuits, offering improved accuracy and efficiency over traditional methods. Ridge regression has been used for aging-aware cell library characterization, reducing guardband settings with accurate degradation estimates \cite{klemme2021_ML_cellLib}. Building on this, an aging-aware path selection method infers delays for a larger set of paths based on measurements from a smaller set, improving critical path delay prediction by accounting for PVT variations \cite{firouzi2013_ML_representative}. In deep neural network (DNN) hardware accelerators, DNN-Life enhances the longevity of weight memory by using memory write and read transducers to measure and mitigate aging, reducing energy consumption \cite{Hanif2021_ML_ageMiti}. Similarly, machine learning-based methods predict near-critical path aging by modeling degradation based on circuit activity and environmental conditions, enabling dynamic performance point selection \cite{kannan2021_ML_RNN_Activityaware}. Reinforcement learning has also been applied to aging mitigation. Life-Guard optimizes task mapping based on aging effects, improving core health and system reliability \cite{Rathore2019_ML_ageMiti}. Deep learning-based methods have further enhanced aging analysis; for example, a fast 2D analysis of electric potential and fields in VLSI chips achieves a 138x speedup over traditional methods while maintaining 99\% accuracy for TDDB aging analysis \cite{Peng2020_ML_TDDB}. Advanced neural networks have been integrated into delay modeling. Aadam combines SPICE simulations with feedforward neural networks (FFNNs) to provide fast and accurate delay predictions, though it may face limitations with diverse cell types \cite{ebrahimipour2020_ML_NN_aadam}. Recurrent neural networks (RNNs) have also been used for aged circuit simulations, incorporating physical constraints to enhance predictive accuracy \cite{Rosenbaum2020_ML_RNN_age}. Machine learning has significantly accelerated FPGA aging analysis. MAPLE predicts FPGA block-level aging 104 to 107 times faster than SPICE with $< 0.7\%$ error, aiding in aging-aware FPGA design \cite{Ghavami2021_ML_DNN_FPGA}. Moreover, real-time machine learning-based aging prediction improves accuracy under dynamic conditions, enabling more effective mitigation strategies \cite{Huang2019_ML_realtime_agepred}.

Few studies have explored ML approaches for predicting and mitigating aging in non-von Neumann architectures such as neuromorphic computing. Neuromorphic systems rely on non-volatile memory for machine learning tasks, but aging of CMOS-based transistors can lead to hardware faults. Aggressive device scaling increases power density and temperature, accelerating aging. Existing de-stressing techniques operate at fixed intervals, causing latency in spike generation and propagation. To address this, an intelligent run-time manager (NCRTM) dynamically de-stresses circuits during machine learning workloads to meet reliability targets, improving hardware reliability with minimal performance impact \cite{song2021a_ML_Neuro_DynamicRM}. Electromigration (EM) sign-off has also become more complex with technological scaling, making modifications to power grid designs time-consuming. A machine learning-based methodology predicts EM-aware aging in power grids using neural network regression and a logistic regression-based classification technique to identify vulnerable metal segments. This model significantly accelerates prediction while achieving higher MTTF values than conventional approaches \cite{Dey2020_ML_EM_miti}.

\subsubsection{Graph Learning-based Cell Aging Prediction and Mitigation}
Applying graph neural networks (GNNs) to predict cell-level aging degradation is promising because the internal topology of a standard cell can be represented as a graph, where transistors serve as nodes and their connections as edges. GNNs can effectively capture the relationship between input conditions and aging effects using an aggregation mechanism to learn neighborhood influences and deep neural layers to model degradation.A temporal-spatial GNN framework combining Graph Temporal Units (GTU) for delay prediction and GraphSAGE for spatial topology analysis was proposed, achieving a 200× improvement in circuit timing estimation efficiency \cite{ye2023_ML_GNN_fast}. A follow-up work later introduced a heterogeneous graph attention network (HGAT) for aging-aware cell timing, enhancing accuracy and efficiency \cite{ye2023_ML_GNN_fast_TCAS2}. Their advanced EdgeGAT network further improved static timing analysis for modern CMOS devices, balancing speed and accuracy \cite{ye2023_ML_GNN_graph}. A two-part approach using graph-attention networks and a path criticality algorithm was also proposed, outperforming classical ML models in identifying aging-critical cells \cite{ye2023_ML_GNN_aging}.

A heterogeneous graph convolutional network (H-GCN) was proposed to estimate aging-induced transistor degradation in analog ICs. The method uses a directed multigraph for IC topology and a probability-based sampling technique for efficient training, outperforming traditional methods in large-scale analog designs \cite{chen2021_ML_GNN_deep}. Another study proposed a GNN-based framework using principal neighborhood aggregation (PNA) for aging-aware static timing analysis (STA), achieving low mean absolute error but limited by its reliance on standard cell types alone \cite{alrahis2022_ML_GNN_gnn4rel}. An HGAT-based method to predict cell delays while managing parasitic RC networks within cells reduced redundant RCs, balancing node imbalances and improving prediction accuracy with a relative RMSE of 2.67\% and a delay mismatch reduction to 1.34 ps—outperforming competitive methods by up to 14.5× in accuracy \cite{cheng2024_ML_GNN_heterogeneous}. Another study proposed a multi-view graph learning framework combining spatial-temporal Transformer networks (STTN) and GAT for aging-aware path-level timing prediction. It achieved a 3.96\% average mean absolute percentage error (MAPE) but struggled with wire delay contributions and increased path-level complexity \cite{bu2024_ML_GNN_multi}. A similar STTN-GAT framework for early-stage STA achieved a 3.75\% MAPE and notable speedup over SPICE simulations, but lacked detailed EDA integration and flip-flop delay modeling \cite{jia2024_ML_GNN_aging}. Table ~\ref{tab:model_comparison_GNN} summarizes the performance of these GNN-based models. Ongoing research explores combining these aging models with dynamic workload balancing to further mitigate aging effects.

\begin{table*}[t!]
\centering
\caption{Comparison of different graph learning-based models and their characteristics.}
\label{tab:model_comparison_GNN}
\setlength{\tabcolsep}{5pt}
\renewcommand{\arraystretch}{1.5}
\resizebox{0.7\textwidth}{!}{
\begin{tabular}{c | c c| c  c| c c c | c }
\toprule
\multirow{2}*{\textbf{Model}} & 
\multicolumn{2}{|c|}{\textbf{Prediction Level}} & 
\multicolumn{2}{c|}{\textbf{Graph Representation}} & 
\multicolumn{3}{c|}{\textbf{Aging Workload}} & 
\multirow{2}*{\textbf{\makecell{Layout \\ Parasitic} }} \\
\cmidrule{2-8}
 & \textbf{Cell} & \textbf{Path} & \textbf{Heter.} & \textbf{Homo.} & \textbf{Temp.} & \textbf{Vgs} & \textbf{Vds} &  \\
\midrule
GraphSAGE + GTU ~\cite{ye2023_ML_GNN_fast}& $\checkmark$ &  & $\checkmark$ & $\checkmark$ & $\checkmark$ & $\checkmark$ & $\checkmark$ & $\checkmark$ \\ 
H-GCN ~\cite{chen2021_ML_GNN_deep}& $\checkmark$ &  & $\checkmark$ &  &  &  & $\checkmark$ &  \\ 
H-GAT ~\cite{ye2023_ML_GNN_fast_TCAS2}& $\checkmark$ &  & $\checkmark$ &  & $\checkmark$ & $\checkmark$ &  & $\checkmark$ \\ 
HGAT ~\cite{cheng2024_ML_GNN_heterogeneous}& $\checkmark$ &  & $\checkmark$ &  & $\checkmark$ & $\checkmark$ &  & $\checkmark$ \\ 
STTN+GAT ~\cite{bu2024_ML_GNN_multi}&  & $\checkmark$ &  & $\checkmark$ & $\checkmark$ & $\checkmark$ &  &  \\ 
PNA-GNN ~\cite{alrahis2022_ML_GNN_gnn4rel}&  & $\checkmark$ &  & $\checkmark$ &  &  &  &  \\ 
EdgeGAT ~\cite{ye2023_ML_GNN_graph}&  & $\checkmark$ &  & $\checkmark$ & $\checkmark$ &  &  &  \\ 
\bottomrule
\end{tabular}
}
\end{table*}

\subsection{EDA-based Optimizations}
Traditional EDA algorithms primarily focus on PPA optimization, with limited attention to reliability. However, increasing guardbands and growing performance demands have driven the development of new optimization techniques to address aging issues. This section explores these techniques in cell libraries, routing, gate replacement, and related areas for characterizing and mitigating aging-related degradation.

\subsubsection{Aging-aware Cell-level Optimization}
Several frameworks have been developed to optimize standard cell libraries from an aging perspective, focusing on improving circuit lifetime and reliability. One key approach is timing analysis with NBTI awareness. \cite{wang2007_ML_efficient} introduces a timing analysis framework incorporating an NBTI-aware library to estimate the importance of gates affected by NBTI. It identifies the worst-case signal probability for each critical path, reducing degradation without overly pessimistic analysis. Applied to ISCAS and ITC benchmark circuits at the 65 nm node, the approach shows that safeguarding just 1\% of gates can prevent timing degradation within 10\% over a decade, with minimal computational cost. Building on this, aging-aware logic synthesis methods have been explored to further extend circuit lifetime. \cite{ebrahimi2013_ML_logicsynthesisaging} proposes a method that optimizes post-aging delays, ensuring paths meet assigned guardbands simultaneously. Tighter timing constraints are applied to highly aged paths, while looser ones are set for less affected paths. Integrated with a commercial synthesis toolchain, this method boosts circuit lifetime by over 3x with minimal area impact.

The development of degradation-aware libraries has enhanced the accuracy of guardband prediction and mitigation. \cite{amrouch2016_ageMiti_Guardband_approx} highlights that incorporating aging effects during logic synthesis increases circuit lifetime. Similarly, \cite{Amrouch2018_ML_aging_aware_opti} presents degradation-aware cell libraries that model BTI's impact on delay, power, and transistor parameters like carrier mobility and gate capacitance. It demonstrates that relying solely on threshold voltage underestimates BTI effects, showing the importance of comprehensive modeling. Advanced optimization algorithms have also been introduced to handle aging-induced degradation. \cite{Kalluru2021_ML_age_aware_opti} introduces Glowworm Swarm Optimization (GSO) and Neighbourhood Cultivation Genetic Algorithm (NSGA) to explore the circuit design space, balancing performance and power under process variations and aging effects. This method reduces critical path latency by 50\% while maintaining the power budget within limits.

Targeting specific components of circuit aging, innovative path isolation and gate replacement techniques have emerged. \cite{Lu2018_ML_ageaware_pathIso} proposes an age-aware synthesis algorithm that isolates aging-sensitive paths, conserving up to 67.7\% of area compared to traditional over-design methods. Similarly, \cite{Qingwu2017_ML_ageaware_opt_gaterepla} introduces a metric for identifying critical gates affected by NBTI and an improved gate replacement algorithm, improving average delay rate by 25.11\% with minimal overhead. Finally, direct circuit-level adjustments have been explored through aging-aware timing analysis. \cite{Mishra2019_ML_aging_aware_14nm_digi} evaluates NBTI effects on 14-nm FinFET-based digital logic, showing that non-uniform library cell extensions can increase circuit lifetime by ~150\% with minimal area overhead. \cite{wu2009_ageMiti_gateSizing, wu2011_ageMiti_gateSizing} develops a timing analysis technique for aging-aware optimization of critical and near-critical sensitizable paths. Using timed automated test pattern generation (ATPG), it identifies sensitizable routes and applies logic reconstruction, pin reordering, transistor scaling, and route sensitization to reduce BTI-induced circuit delay.

These advancements in aging-aware standard cell libraries, logic synthesis, path isolation, and gate replacement have collectively improved the reliability and lifetime of integrated circuits under aging effects, highlighting the growing importance of reliability-driven EDA optimization.

\subsubsection{Age-aware Routing}
The Network-on-Chip (NoC) is a crucial solution for managing communication among multiple cores in multi-core system-on-chip (SoC) architectures. As feature sizes shrink, reliability has become a major challenge due to aging effects. Uneven load conditions and traffic within the NoC create hotspots, particularly at core routers, which accelerate aging-induced degradation. To address this, several aging-aware routing and optimization strategies have been proposed. One approach is the Location-based Aging-resilient Xy-Yx (LAXY) routing algorithm, which redistributes packet flow to balance traffic across core nodes and reduce stress. Simulations show that the Fishtail configuration improves the mean time to failure (MTTF) of routers and interconnects by 42\% and 56\%, respectively, while reducing packet delay by 7\% with minimal area overhead \cite{Rohbani2017_ML_ageaware_NOCRoute}.

For 3D NoC-based chip multiprocessors (CMPs), aging is further exacerbated by gate-delay degradation and electromigration. To mitigate this, a runtime framework (ARTEMIS) has been proposed for dynamic application mapping and voltage scaling. ARTEMIS regulates aging and power distribution, enhancing chip lifespan and addressing dark-silicon constraints. Experimental results show that ARTEMIS supports 25\% more applications over the chip’s lifetime compared to prior methods \cite{Raparti2017_ML_ageaware_NOCRoute}. A more comprehensive solution is RETUNES, a five-tier voltage and frequency scaling methodology that balances power savings and reliability under varying network loads. RETUNES combines near-threshold voltage (NTV) scaling with an adaptive routing algorithm to reduce power consumption and improve network performance. It achieves approximately 2.5$\times$ higher power savings and a threefold improvement in the energy-delay product for Splash-2 and PARSEC benchmarks \cite{Bhamidipati2019_ML_ageaware_NOCRoute}. SCRA, a deterministic aging-resilient hybrid routing algorithm, distributes packet flow uniformly across the network to mitigate aging. It employs a flow distribution model to enhance network longevity and communication performance compared to one-dimensional order routing methods \cite{Zhang2019_ML_ageaware_NOCRoute}. An online monitoring and adaptive routing framework has also been proposed using a Centralized Aging Table (CAT). CAT tracks router aging under different stress and temperature ranges and dynamically adjusts routing paths to avoid highly aged routers. Simulation results using the gem5 simulation flow demonstrate that this method reduces maximum router aging and aging imbalance by 39\% and 52\%, respectively, with minimal impact on network latency and energy-delay product \cite{Ghaderi2017_ML_ageaware_NOCRoute}.

In FPGA-based NoC designs, aging impacts routing reliability and signal integrity. A routing technique has been proposed to balance stress across routing resources, reducing buffer deterioration and improving signal integrity. This strategy reduces timing guardbands by 14.1\% to 31.7\%, depending on the FPGA routing architecture \cite{Khaleghi2019_ML_ageaware_NOCRoute}.

These strategies highlight the importance of aging-aware routing and optimization in NoC designs, offering significant improvements in system reliability, lifetime, and performance under aging-induced degradation.

\subsection{Approximate Computing for Aging Mitigation}
Approximate computing is a promising design paradigm that enhances system efficiency by leveraging error tolerance in applications like image processing and neural networks, striking a balance between implementation complexity and computational accuracy \cite{Leon2025_ApproximateComputing}. It has gained increasing attention for reducing critical path delays in circuits, ensuring maximum frequency operation throughout their lifetime. Adaptive approximate computing has emerged as an alternative to traditional guardband techniques for aging mitigation. Early work focused on minimizing aging-related issues via approximate computing in critical delay paths, logic gates, and cell libraries \cite{amrouch2016_ageMiti_Guardband_approx, Amrouch2017_ageMiti_dig_approx}. One approach employs adaptive approximations to trade off transient, degradation-induced variations in circuit delays for permanent performance improvements with minimal quality loss. This method achieves up to 21\% speedup with less than 2\% area and energy impact compared to traditional guardbanding \cite{Boroujerdian2018_ML_approx}. A novel design methodology converts aging-induced timing violations into computational approximation errors without increasing supply voltage. This approach enhances image quality while reducing dynamic and static power consumption by 21.45\% and 10.78\%, respectively, with only 0.8\% area overhead \cite{Kim2020_Approx_ageMiti}. An automated framework for aging-aware circuit approximation has also been proposed, using directed gate-level netlist approximation to introduce small functional errors and recover from delay degradation. The framework reduces error by 1208 times compared to baseline circuits \cite{Balaskas2021_ML_approxComp}.

Approximate computing has also been applied to nanoscale semiconductor technologies to mitigate aging-related degradation under high temperatures. A study explored various approximate arithmetic circuits, such as adders and multipliers, using advanced approximation techniques. Results show that truncated arithmetic circuits result in higher quality loss, but a truncated multiplier achieves minimal error distance, enabling guardband reduction while maintaining high visual quality in image processing applications \cite{Santiago_2022_ML_approxCompu}.

These advances demonstrate that approximate computing effectively balances performance, power, and accuracy while addressing aging-related reliability issues, making it a valuable strategy for long-term circuit optimization.

\subsection{Aging Characterization and Mitigation on Physical Unclonable Functions (PUFs)}
Physical Unclonable Functions (PUFs) are hardware-based security primitives that exploit inherent manufacturing variations in integrated circuits (ICs) to generate unique, unclonable identifiers for authentication and cryptographic key generation. However, aging mechanisms such as BTI, HCI, TDDB degrade PUF reliability over time, compromising long-term security \cite{karimi2018_ML_PUF_ImpactAgingReliability}. This section examines recent advancements in aging-resistant PUF designs.

A current-starved inverter structure with four MOSFETs has been proposed to improve ring-oscillator (RO) PUF reliability. Two additional transistors control the drain current, enhancing stability compared to conventional inverter-based ROs \cite{liu2017_ML_PUF_ACROPUFLowpowerReliable}. Another study introduces an aging-resilient RO PUF for FPGA hardware, addressing transient variations and aging effects using SRAM cells and multiple paths in FPGA look-up tables (LUTs). This design increases reliability by 37.4\% and reduces aging degradation by 37\% compared to traditional FPGA-based RO PUFs, without requiring circuit-level redesign \cite{chowdhury2017_ML_PUF_AgingResilientRO}. An aging-resistant PUF based on commercial NAND flash memory employs a “program-disturb” technique and an adaptively tunable approach to mitigate aging effects. This method ensures consistent randomness and reliability over at least 1000 PUF-generating operations \cite{sakib2020_ML_PUF_AgingResistantNANDFlash}.

The Highly BTI Resilient RO (HBTIRRO) PUF demonstrates enhanced resistance to BTI degradation. It offers two implementations with different trade-offs in resilience, energy consumption, and area. HBTIRRO reduces frequency degradation by 57.2\% and 34.6\%, respectively, while maintaining comparable area and slightly higher power consumption. Its improved robustness and reduced area overhead make it a secure and efficient solution for RO-based PUF designs \cite{omana2024_ML_PUF_AgingResilientRing}.

These advancements highlight the potential of design-level techniques and architectural innovations to mitigate aging effects, improving the long-term reliability and security of PUF-based hardware authentication.

\begin{table*}[!t]
\centering
\caption{Current challenges and future directions of IC reliability.}
\label{tab:challenges_future_directions}
\setlength{\tabcolsep}{5pt}
\renewcommand{\arraystretch}{1.25}
\resizebox{0.9\textwidth}{!}{
\begin{tabular}{p{6.8cm}|p{7.5cm}}
\toprule
\textbf{Current Challenges} & \textbf{Future Directions} \\ 
\midrule

\textbf{Physics-based Aging Models:} Computationally expensive and challenging to modeled aging mechanisms (BTI, HCI, TDDB, and EM) at sub-10nm having high current densities and 3D structures. 

& \textbf{ML-based models:} Development of ML algorithms (DNN, RNN, GNN etc.,) for predicting real-time aging prediction and preemptive system-level adjustments. \\ 

\textbf{Stochastic Behaviors:} Increased computational complexity for the development of universal statistical models that combines both process and aging-induced variations 

 & \textbf{Advanced Materials and Cooling:} Innovative materials (e.g., high-k, graphene) and advanced packaging (e.g., 3DICs with microfluidic cooling \cite{Wang2024_cool3DIC}) can improve device longevity and thermal stability \\ 

\textbf{Trade-off in Traditional Mitigation Techniques:} Guardbanding compromises performance and energy efficiency. Adaptive techniques require real-time monitoring which increases design complexity. 

& \textbf{Hardware-software Co-design:} Unified frameworks and reconfigurable architectures can optimize task allocations and can dynamically reroute to reduce stress on critical components. \\ 

\textbf{On-chip Aging Sensors:} These are vulnerable to noise, jitter, and accelerated aging, which complicating calibration and eventually lead to inaccurate data ~\cite{Wang2025_Review_testingMethods}. 

& \textbf{Standardization and Collaboration:} Interdisciplinary collaboration for innovative materials, and standardization of aging models across all technology nodes by collaborating academia and industry. EDA tools should integrate predictive aging models into the design flow. \\ 

\bottomrule
\end{tabular}
}
\end{table*}
\section{Challenges and Future Directions}
\label{challenge}
As CMOS technologies advances to smaller nodes, ICs encounter increasing reliability issues stemming from complex interactions among aging mechanisms, including BTI, HCI, TDDB and EM. Despite developments in modeling and mitigation strategies for reliability management, persistent challenges, such as computational inefficiencies in physics-based models, process variability, and the trade-offs between guardbanding and performance—necessitate novel solutions. Future advancement depends on interdisciplinary approaches, including machine learning (ML)-based predictive modeling, advanced materials (e.g., high-k dielectrics), and system-level techniques like as hardware-software co-design. Table ~\ref{tab:challenges_future_directions} highlights these challenges and suggests focused research directions to achieve resilient, aging-aware integrated circuit design.

\section{Conclusions}
\label{conclusions}
Driven by increased reliability challenges from shrinking device dimensions and rising demands for reliable chips in diverse emerging applications, this survey explores the complex issue of aging in integrated circuits, covering both fundamental mechanisms and advanced mitigation techniques. It examines key degradation factors, including Bias Temperature Instability (BTI), Hot Carrier Injection (HCI), Time-Dependent Dielectric Breakdown (TDDB), Electromigration (EM), and stochastic aging-induced variations, which collectively degrade the reliability and performance of modern ICs. Primary aging monitoring methods, such as ring-oscillator-based techniques, critical path replicas, and aging sensor circuits, are discussed for both offline and real-time degradation measurement. Aging mitigation strategies at both circuit and system levels are analyzed in detail. Circuit-level approaches address the distinct aging characteristics of digital, analog, and SRAM circuits, while system-level techniques complement these solutions by managing interactions among components in modern electronic systems. The survey also highlights emerging methods leveraging machine learning, graph-based learning, approximate computing, and aging-aware cell libraries, offering potential solutions to the increasing complexity of aging management in advanced technology nodes.

Despite significant advancements in IC design, key challenges remain, including the need for more accurate aging models, better handling of process variability and environmental influences, and improved integration of monitoring and mitigation strategies throughout design stages. Future research should focus on developing standardized benchmarking frameworks, creating extensive datasets for machine learning applications, and exploring innovative materials and device architectures that inherently mitigate aging effects. As the semiconductor industry pushes toward greater miniaturization and performance, addressing these aging-related challenges will be critical for ensuring the reliability and longevity of integrated circuits in the era of sustainable and reliable computing.





\ifCLASSOPTIONcaptionsoff
  \newpage
\fi

\bibliographystyle{IEEEtran}
\balance
\bibliography{ms}
\vspace{-30pt}


\end{document}